\documentclass[pra,aps,twocolumn,showpacs,superscriptaddress]{revtex4}
\usepackage{graphicx}
\usepackage{multirow}
\usepackage{amsmath}
\usepackage{amssymb}
\usepackage{amsthm}
\usepackage{eucal}
\usepackage{bm}
\usepackage{upgreek}
\usepackage{mathrsfs}
\usepackage{latexsym}
\usepackage{float}
\usepackage{hyperref}
\usepackage{setspace}

\theoremstyle{plain}
\newtheorem{thm}{Theorem}
\theoremstyle{plain}
\newtheorem{theorem}[thm]{Theorem}
\newtheorem{proposition}[thm]{Proposition}
\newtheorem{conjecture}[thm]{Conjecture}
\newtheorem{corollary}[thm]{Corollary}

\newtheorem{remark}[thm]{Remark}

\newcommand{\ket} [1] {\vert #1 \rangle}
\newcommand{\bra} [1] {\langle #1 \vert}
\newcommand{\braket}[2]{\langle #1 | #2 \rangle}
\newcommand{\proj}[1]{\ket{#1}\bra{#1}}

\newcommand{\hr}[1]{\hat{\rho}_{#1}}
\newcommand{\bs}[1]{\boldsymbol{#1}}
\newcommand{\D}[1]{{\cal D}(\boldsymbol{#1})}

\begin{document}

\singlespacing

\title{Non-Gaussian operations on Bosonic modes of light: \protect\\ photon-added Gaussian channels}

\author{Krishna Kumar Sabapathy}
\email{krishnakumar.sabapathy@gmail.com}
\affiliation{F\'{i}sica Te\`{o}rica: Informaci\'{o} i Fen\`{o}mens Qu\`{a}ntics, Departament de F\'{i}sica, Universitat Aut\`{o}noma de Barcelona, ES-08193 Bellaterra (Barcelona), Spain.}

\author{Andreas Winter}
\email{andreas.winter@uab.cat}
\affiliation{F\'{i}sica Te\`{o}rica: Informaci\'{o} i Fen\`{o}mens Qu\`{a}ntics, Departament de F\'{i}sica, Universitat Aut\`{o}noma de Barcelona, ES-08193 Bellaterra (Barcelona), Spain.}
\affiliation{ICREA---Instituci\'{o} Catalana de Recerca i Estudis Avan\c{c}ats, Pg.~Lluis Companys, 23, ES-08001 Barcelona, Spain.}

\begin{abstract}
We present a framework for studying Bosonic non-Gaussian channels of continuous-variable systems. Our emphasis is on a class of channels that we call photon-added Gaussian channels which are experimentally viable with current quantum-optical technologies. A strong motivation for considering these channels is the fact that it is compulsory to go beyond the Gaussian domain for numerous tasks in continuous-variable quantum information processing like entanglement distillation from Gaussian states and universal quantum computation. The single-mode photon-added channels we consider are obtained by using two-mode beamsplitters and squeeze operators with photon addition applied to the ancilla ports giving rise to families of non-Gaussian channels.  For each such channel, we derive its operator-sum representation, indispensable in the present context. We observe that these channels are Fock-preserving (coherence non-generating). We then report two novel examples of activation using our scheme of photon addition, that of quantum-optical nonclassicality at outputs of channels that would otherwise output only classical states, and of both the quantum and private communication capacities, hinting at far-reaching applications for quantum-optical communication. Further, we see that noisy Gaussian channels can be expressed as a convex mixture of these non-Gaussian channels.  We also present other physical and information-theoretic properties of these channels. 
\end{abstract}


\maketitle

\section{Introduction}
Non-Gaussian states and operations have recently received much attention with respect to theoretical and experimental schemes in continuous-variable  quantum information theory. Experimentally realizable non-Gaussian states include Fock states \cite{fock1,fock2,motes}, noon states \cite{noon}, cat states \cite{suzuki,cat}, photon-added coherent states \cite{agarwal,bellini04}, and photon-added thermal states \cite{pats1,pats2}, among other examples \cite{grangier04,polzik06,grangier12}. 
Commonly used non-Gaussian operations include photon addition \cite{bellini07,fiurasek09}, photon subtraction \cite{chefles,psub,sasaki07}, photon counting \cite{pcoun}, cubic phase gates\cite{miyata}, and Kerr nonlinearities \cite{kerr}. 

There are various motivations and uses for going beyond the Gaussian regime with regard to implementation of  quantum information protocols. These include no-go theorems against a Gaussian-only toolbox like distillation of entanglement from Gaussian states \cite{nogo1,nogo2,nogo3}, use as quantum repeaters \cite{guha14}, and for other quantum information protocols like cloning \cite{wolf}, error-correction \cite{cerf09}, bit-commitment \cite{cerf10}, computing with cluster states \cite{eisert10}, and universal quantum computation \cite{lloyd,sanders,nielsen}.  Also non-Gaussian resources have proven advantageous in many scenarios like parameter estimation \cite{adesso}, generation of entangled states \cite{grangier07,sasaki,robust,raul12}, and teleportation \cite{opat,bonifacio,mista06,anno07}.   

\begin{figure}
\centering 
\includegraphics[width=\columnwidth]{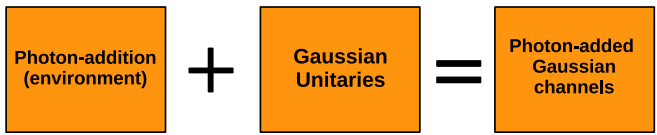}
\caption{Schematic diagram for the construction of a class of non-Gaussian channels using two constituent elements of photon addition and Gaussian unitary gates.}
\label{fig1}
\end{figure}

In this article we generate non-Gaussian operations using two main ingredients, the commonly used (experimentally viable) photon addition and the ubiquitous class of Gaussian unitaries, specifically two-mode squeezers and beamsplitters. We call the resulting non-Gaussian operations as photon-added Gaussian channels and this is schematically represented in Fig. \ref{fig1}. 
Here photon addition is  applied to the environment state in the Stinespring dilation of the underlying Gaussian channel. As a consequence we generate non-Gaussian operations on the initial system when the environment system is ignored.

The method can also be thought of as being one example of the many protocols and implementations which concern manipulating the environment state in the Stinespring representation of a channel that have been considered in literature. Some illustrative examples include implementation of general gates \cite{barry}, using mixed environment states for channel simulation \cite{smolin}, manipulating the environment to generate additional capacities either as a helper or adversary \cite{sid1,sid2,andi.env}, and using feedback from the environment to correct for transmission of information through the channel \cite{grego1,grego2,king}. 

We stress that although we use the common terminology  `environment system' to denote the additional system in the Stinespring representation, we think of it as auxiliary ports one can keep track of as inputs to unitary gates used in the implementation. Hence  the channels that we introduce and study are implementable non-Gaussian gadgets of potential use in continuous-variable quantum information processing. 

The rest of the article is structured as follows. Section II introduces the basic framework for quantum channels, the phase-space techniques, and the phase-space description of photon-added channels. In Section III we present our construction of photon-added channels and obtain the operator-sum representation of three special classes of channels closely related to attenuation, amplification, and phase conjugation channels. We then study in Section IV numerous physical, mathematical, and information-theoretic properties of the photon-added channels introduced in Section III, and conclude in Section V. 

\section{Preliminaries \label{prelim}}
We now discuss the connection between the unitary Stinespring representation and the operator-sum representation of any quantum channel. Given a channel $\Phi$ it can always be represented as a unitary dynamics on a suitably chosen larger system with the ancilla system consequently ignored. Let the input system be described by a Hilbert space $\mathcal{H}_A$. Then there exists an environment with Hilbert space $\mathcal{H}_E$, a system-environment unitary $U_{AE}$, and a pure state $\ket{\psi}_E \in {\cal H}_E$, such that any completely positive trace-preserving map on system $A$ can be represented in the following way:
\begin{align}
\Phi[\hr{A}] = {\rm Tr}_E\, \left[ U_{AE}\,(\hr{A} \otimes \ket{\psi}_E \bra{\psi})\,U_{AE}^{\dagger} \right].
\label{e1}
\end{align}
To obtain an operator-sum representation \cite{kraus} we introduce a suitable complete basis to perform the trace of the environment system. Since our physical system is that of Bosonic modes of an electromagnetic field or harmonic oscillators, we  make use of the Fock basis. Choosing the Fock basis $\{\ket{k}_E \}$, we have from Eq. \eqref{e1}, 
\begin{align}
\Phi[\hr{A}] &= \sum_k\,  {}_E \bra{k} \,\left[ U_{AE}\,(\hr{A} \otimes \ket{\psi}_E \bra{\psi})\,U_{AE}^{\dagger} \right] \ket{k}_E \nonumber\\
&= \sum_k\,  ({}_E \bra{k} \, U_{AE} \,\ket{\psi}_E) \,\hr{A} \,({}_E \bra{\psi} \,U_{AE}^{\dagger}\, \ket{k}_E)  \nonumber\\
&= \sum_k\, F_k \,\hr{A} \, F_k^{\dagger},  
\label{e2}
\end{align}
where $\left\{ F_k = {}_E \bra{k}  U_{AE} \ket{\psi}_E \right\}$ are the Kraus operators for the channel $\Phi$.   

We first consider the case when the system and the environment are both single-mode systems. The corresponding generalization to the case of multimodes can be performed in a similar way. Let the two-mode unitary operator in the dilation of a channel $\Phi$  be written as $U[\Phi]$ with matrix elements in the Fock basis given by 
\begin{align}
U[\Phi] = \sum_{m_1,m_2,n_1,n_2=0}^{\infty} T[\Phi]_{n_1,n_2}^{m_1,m_2} \ket{m_1,m_2} \bra{n_1,n_2}. 
\end{align}
Then by  Eq. \eqref{e2} we have that the Kraus operators $\{ F_{k} \}$ of $\Phi$
are given by 
\begin{align}
F_{k} &= {}_E\bra{k} \left[ \sum_{m_1,m_2,n_1,n_2=0}^{\infty} \hspace{-0.6cm} T[\Phi]_{n_1,n_2}^{m_1,m_2} \ket{m_1,m_2} \bra{n_1,n_2}\right] \ket{\psi}_E\nonumber\\
&= \sum_{m_1,n_1,n_2=0}^{\infty} T[\Phi]_{n_1,n_2}^{m_1,k} \, \braket{n_2}{\psi}_E \,\ket{m_1} \bra{n_1}.
\label{kraus-rep}
\end{align}

In this article we are interested in channels for which there exists a dilation $ U_{AE}$ that is Gaussian even if the environment state $\ket{\psi}_E$ is not necessarily Gaussian. We call such channels {\em Gaussian dilatable}. 
\begin{figure}
\centering
\scalebox{0.55}{\includegraphics{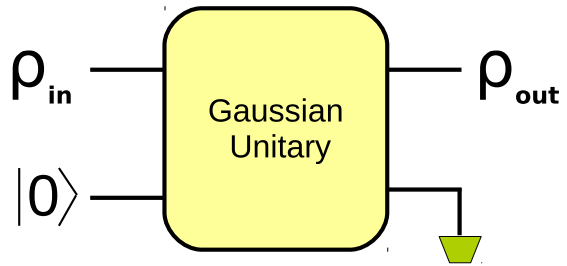}}
\caption{Unitary dilation for Gaussian channels. Any quantum-limited single-mode Bosonic Gaussian channel can be realized in the following three step method: first one appends a suitable single-mode environment state which can be taken to be pure Gaussian (vacuum, without loss of generality), then evolves the system-environment state through a canonical two-mode Gaussian unitary, and finally, traces out the environment degrees of freedom.}
\label{fig2}
\end{figure} 
We first recall a few points regarding the Stinespring dilation of Gaussian channels since it is our starting point as hinted in Fig. \ref{fig1}. We first focus on the case of quantum-limited channels. For the underlying  phase-space framework we refer the reader to \cite{nb,nbm}. An important fact that will play a crucial role for us is that every single-mode quantum-limited Gaussian channel can be realized as a canonical unitary interaction with a suitable single-mode environment taken in a pure Gaussian state (or vacuum state without loss of generality) and then ignoring the environment degrees of freedom. This representation is schematically depicted in Fig. \ref{fig2}. We point out that for noisy single-mode Gaussian channels one can follow a similar method but the main difference being that one requires a two-mode pure environment state. We will make further comments regarding the case of noisy Gaussian channels towards the end of the article in Section \ref{noisypa}.

\subsection{Phase-space description of linear Bosonic channels}
\begin{figure*}
\centering
\scalebox{0.55}{
\includegraphics{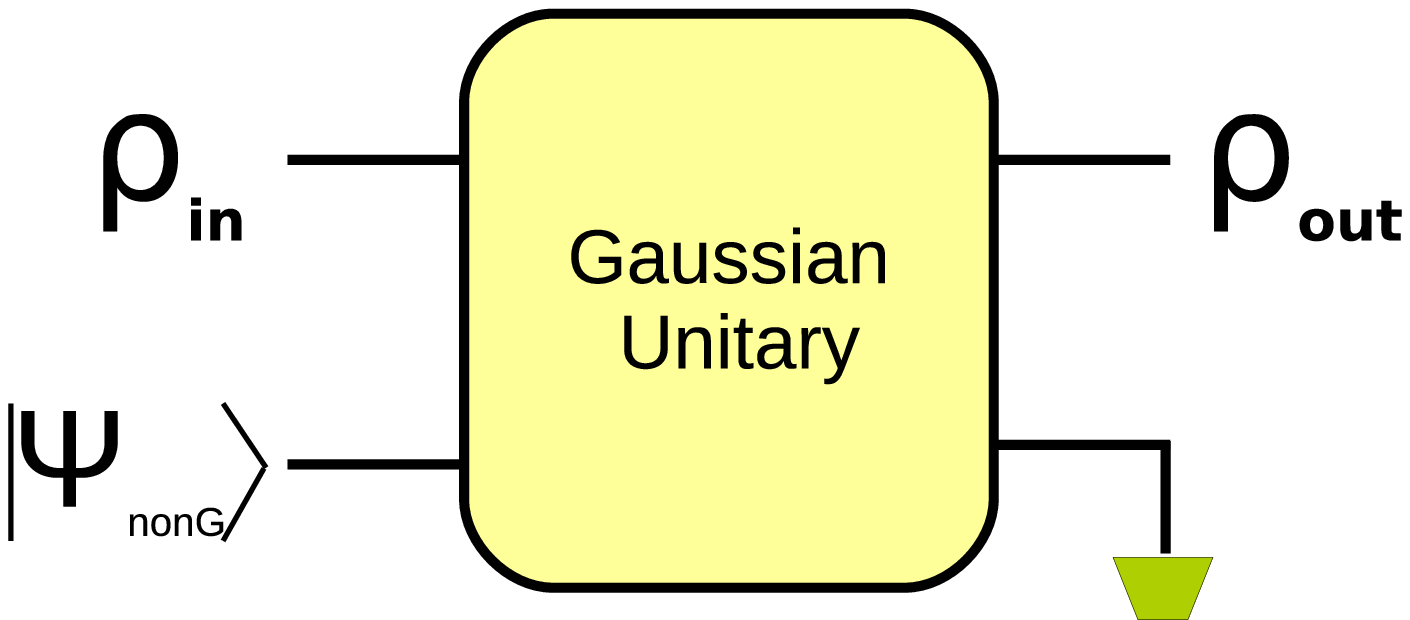}
\hspace{2cm}
\includegraphics{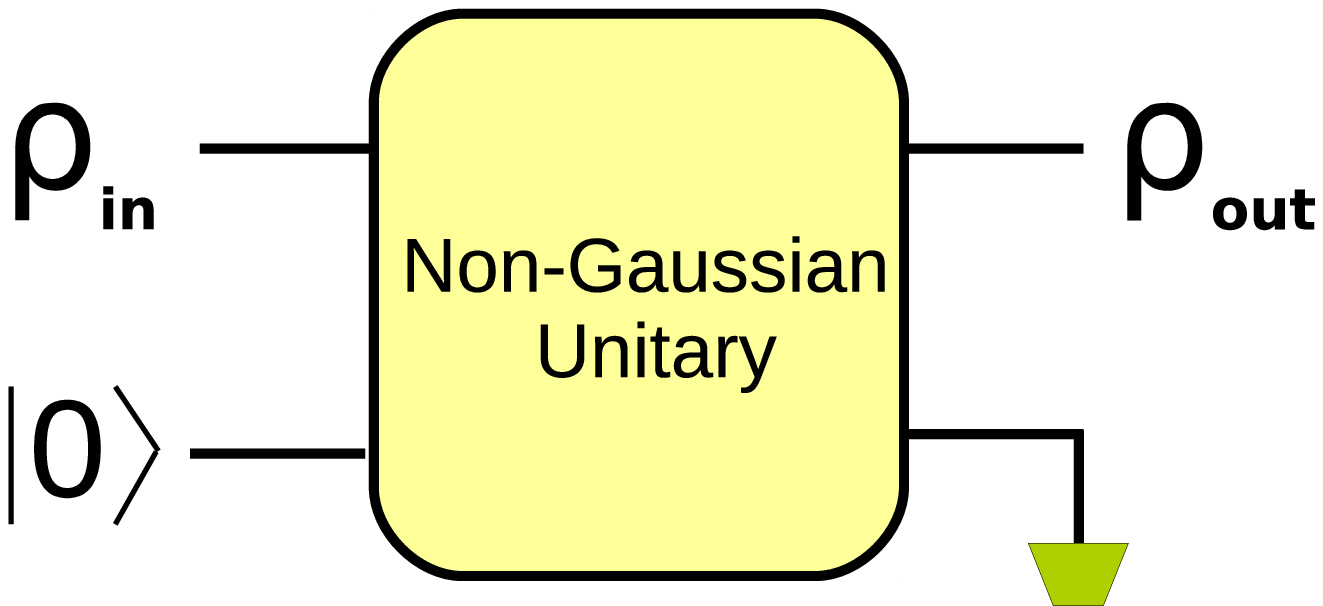}}
\caption{Basic schemes to construct non-Gaussian channels using the Stinespring representation. One can use an ancilla non-Gaussian state $\ket{\psi_{\rm nonG}}$ and a Gaussian unitary as in scheme 1 (left) and we label this scheme as Gaussian dilation scheme. The second option is to use a non-Gaussian unitary with a Gaussian ancilla as in scheme 2 (right) which we label as the general dilation scheme. A third scheme that combines the use of a non-Gaussian unitary and a non-Gaussian ancilla state can be reduced to scheme 2. In this article we  focus on scheme 1. \label{fig3}}
\end{figure*} 
Non-Gaussian channels can be realized by two underlying schemes in its Stinespring dilation as depicted in Fig. \ref{fig3}. The first scheme is when  the non-Gaussian channel is realized through a Gaussian unitary and a non-Gaussian environment state which we call the Gaussian dilation scheme. The second scheme is when we use  a non-Gaussian unitary and a Gaussian environment state and we call this the general dilation scheme. The final  combination of a non-Gaussian unitary with a non-Gaussian environment state can be reduced to the general dilation scheme. For simplicity we consider the system and environment to be a single-mode system both from a theoretical and experimental point of view. The generalization to the multimode case can be done in an analogous manner. 

We focus on the first Gaussian dilation scheme where we have a Gaussian unitary and a non-Gaussian environment state. In particular, as mentioned earlier, we construct a special class of channels that we call as photon-added Gaussian channels. These channels are constructed through three elementary building blocks. First we start with the Gaussian unitary in the dilation of a Gaussian channel and a vacuum state in the environment as previously depicted in Fig. \ref{fig2}. Next we apply a suitable photon addition to the environment state and finally, we trace out the environment system. We now briefly discuss the phase-space description of photon-added Gaussian channels.  

Let the Gaussian unitary of the dilation corresponding to the underlying Gaussian channel induce a symplectic transformation of the form 
\begin{align}
S^{-1} = \left[ \begin{matrix}
S_{11} & S_{12} \\
S_{21} & S_{22}
\end{matrix} \right]
\end{align}
at the level of the characteristic function \cite{simon00}, i.e.,
\begin{align}
\chi(\bs{\xi};U[S]\,\hat{\sigma} \,U[S]^{\dagger}) = \chi(S^{-1} \bs{\xi};\hat{\sigma}). 
\label{e6}
\end{align}
Let $\hr{E}$ be the photon-added state of the environment. So we have by  Eqs. \eqref{e1} and \eqref{e6} that the combined system-environment output state after the action of the Gaussian unitary is 
\begin{align}
&\chi({\bs{\xi}_1},{\bs{\xi}_2}; U[S] (\hr{} \otimes \hr{E}) U[S]^{\dagger}) \nonumber\\
&=  
\chi(S_{11} \bs{\xi}_1 + S_{12} \bs{\xi}_2;\hr{}) \,\chi(S_{21} \bs{\xi}_1 + S_{22} \bs{\xi}_2;\hr{E}).
\label{e7}
\end{align}
We then obtain the characteristic function of the output state of the channel corresponding to input state $\hr{}$ by setting $\bs{\xi}_2 =0$ in  Eq. \eqref{e7}, i.e., 
\begin{align}
\chi(\bs{\xi}_1; \hr{\rm out}) = \chi(S_{11} \bs{\xi}_1;\hr{}) \,\chi(S_{21} \bs{\xi}_1;\hr{E}).
\label{e8}
\end{align}

We now briefly outline the phase space description of linear Bosonic channels of an $n$-mode system. Let $\Phi$ denote a quantum channel and $\Phi^{\dagger}$ the adjoint of $\Phi$. Let $\hr{}$ be an input state to $\Phi$ and let $\Phi(\hr{}) = \hr{}^{\prime}$ be the corresponding output state. At the level of displacement operators ${\cal D}(\bs{\xi})$, or equivalently the Heisenberg picture, $\Phi$ is said to be a linear channel if it can be written as \cite{hw01} 
\begin{align}
\Phi^{\dagger} [\D{\xi}] = {\cal D}(X \bs{\xi}) f(\bs{\xi}), 
\label{e3}
\end{align}
where $X$ is a linear map on the phase space variables $\bs{\xi}$ and $f$ is an arbitrary function which together with $X$  has to satisfy certain constraints encoding the fact that $\Phi$ is completely positive \cite{hw01}. At the level of the characteristic function we  have by Eq.  \eqref{e3} the action of $\Phi$ to be
\begin{align} 
\chi(\bs{\xi};\hr{}^{\prime}) = {\rm Tr} \left[ \Phi(\hr{})\, \D{\xi} \right] &= {\rm Tr} \left[ \hr{}\,  \Phi^{\dagger}(\D{\xi}) \right]\nonumber\\ 
&= {\rm Tr} \left[ \hr{}  {\cal D}(X \bs{\xi}) \right] f(\bs{\xi})\nonumber\\
&= \chi(X \bs{\xi};\hr{}) f(\bs{\xi}).
\label{e4}
\end{align}
When the function $f$ is Gaussian and satisfies the complete-positivity condition, the map corresponds to a linear Bosonic Gaussian channel. For a non-Gaussian function $f$ that satisfies the complete-positivity condition the corresponding channel is a linear Bosonic non-Gaussian channel.

Comparing  Eqs. \eqref{e8} and \eqref{e4} we see that $S_{11}$ plays the role of $X$ and $\chi(S_{21} \bs{\xi}_1;\hr{E})$ plays the role of $f(\bs{\xi}_1)$. Note that we did not make use of any special form for the environment state. Since the map is completely positive trace preserving by construction, we have the following proposition. 
\begin{proposition}
All Gaussian dilatable channels, including the class of photon-added Bosonic Gaussian channels, belong to the class of linear Bosonic channels. 
\end{proposition}

 We now state in the form of a conjecture a connection between the set of Gaussian dilatable channels and linear Bosonic channels. 
\begin{conjecture}
The set of linear Bosonic channels and the set of Gaussian dilatable channels are identical. 
\end{conjecture}
From the proposition mentioned above, what remains to be shown is the converse statement that the set of linear Bosonic channels (i.e., linear Gaussian and linear non-Gaussian) are Gaussian dilatable. 

\begin{table}
\caption{Notation used for the various quantities. ${\cal A}(\kappa;n)$, $\kappa \geq 1$; 
${\cal B}(\kappa;n),$, $\kappa \leq 1$, and ${\cal C}(\kappa;n)$, $\kappa \geq 0$ denote the photon-added amplifier, attenuator, and phase-conjugation channels, respectively. The argument $\kappa$ stands for the corresponding underlying quantum-limited Gaussian channel parameter and $n$ for the resulting number state of the environment post photon addition. $U$ denotes the unitary matrix in the dilation of the underlying Gaussian channels and $T$ represents its Fock basis matrix elements. }
\begin{tabular}{cccc}
\hline
Channel &  Unitary & Kraus & Fock matrix elements   \\
$\Phi$& $U[\Phi]$& operators& of $U[\Phi]$\\
\hline
\hline 
${\cal A}(\kappa;n)$  & $U[{\cal A}(\kappa;n)]$ &$\{A_{\ell}(\kappa;n) \}$ & $T[{\cal A}(\kappa;n)]^{m_1,m_2}_{n_1,n_2}$\\
${\cal B}(\kappa;n)$ &$U[{\cal B}(\kappa;n)]$ &$\{B_{\ell}(\kappa;n) \}$ & $T[{\cal B}(\kappa;n)]^{m_1,m_2}_{n_1,n_2}$\\
${\cal C}(\kappa;n)$ &$U[{\cal C}(\kappa;n)]$ & $\{C_{\ell}(\kappa;n) \}$ & $T[{\cal C}(\kappa;n)]^{m_1,m_2}_{n_1,n_2}$\\
\hline
\end{tabular}
\label{table0}
\end{table}

\section{Kraus decomposition for photon-added Gaussian channels}
\begin{figure}
\centering
\includegraphics[width=\columnwidth]{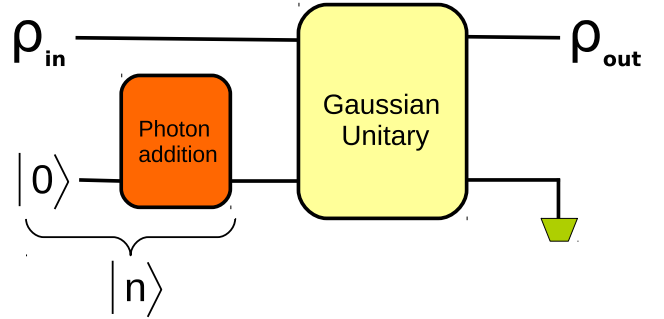}
\caption{Scheme for obtaining photon-added Gaussian channels. Starting from the underlying unitary representation for quantum-limited attenuator, amplifier, and phase conjugation channels we obtain the corresponding photon-added version by applying photon addition to the vacuum state to generate a Fock state in the environment. The resulting channels fall under the class of linear Bosonic non-Gaussian channels. In the absence of photon addition we have the underlying single-mode quantum-limited Bosonic Gaussian channel. \label{fig4}}
\end{figure}

For simplicity, we now consider the case of the photon-added Bosonic Gaussian channels as depicted in Fig. \ref{fig4}, i.e. when the initial environment state is a vacuum state and the underlying Gaussian channel is quantum-limited. 
The case of photon addition to noisy Gaussian channels will be considered towards the end of the article in Sec. \ref{noisypa}. 
We will consider three main classes which are the photon-added quantum-limited amplifier, attenuator, and phase-conjugation channels and study their properties in subsequent sections. It turns out to be useful  to work completely in the Fock basis. For example, computing output states in a certain basis or evaluating entropies and other information theoretic quantities all require working in a specific basis as opposed to a global picture. The reason some of these necessities can be bypassed for the Gaussian case is because many of the quantities are directly computable from the phase space picture which also contains the covariance matrices. 

As mentioned earlier these special illustrative classes have both theoretical importance and are also practically implementable. We will see later that these special classes of single-mode quantum-limited attenuator, amplifier and phase conjugation channels all have a Stinespring dilation where the unitary acting on the larger Hilbert space is either a two-mode beamsplitter or a two-mode squeeze operator. Also, turning on and off this photon addition takes one from the Gaussian to the non-Gaussian regime and hence lends to a comparison of the two schemes in a systematic way. There are other such properties that we explain in the rest of the article. 

Before we begin with our specific examples of non-Gaussian channels we make a brief note of the notation  that we use in the rest of the article and this is outlined in Table. \ref{table0}. We use the calligraphic letters ${\cal A}$, ${\cal B}$, ${\cal C}$ to denote the amplifier, attenuator, and phase-conjugation channels, respectively, with $\kappa$ denoting the channel parameter of the underlying quantum-limited channels and $n$ denoting the Fock state (or number of photon additions applied) of the environment. The regular letters $A$,
$B$, $C$ denote the respective Kraus operators, $U$  the unitary operator of the dilation of the Gaussian channels, and $T$  the matrix elements of this unitary operator in the Fock basis. 

The rest of this section is dedicated to the derivation of an operator-sum representation of the photon-added  quantum-limited amplifier, attenuator, and phase-conjugation channels of a single mode taken in their respective canonical forms.

\subsection{Photon-added amplifier ${\cal A}(\kappa;n)$}
\begin{figure}
\centering
\scalebox{0.75}{\includegraphics{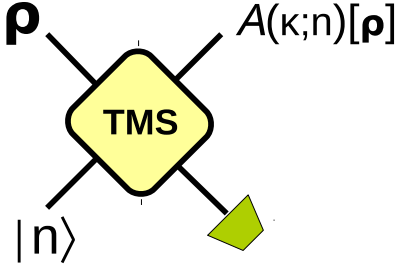}}
\caption{Realization of the photon-added amplifier channels ${\cal A}(\kappa;n)$. Here $\kappa$ denotes the channel parameter coming from the two-mode squeeze (TMS) operator unitary and $n$ denotes the photon addition.\label{fig4a}}
\end{figure} 

We first consider the case of the photon-added quantum-limited amplifier depicted in Fig. \ref{fig4a}. From the scheme mentioned above we have to evaluate the appropriate matrix elements of the unitary operator to obtain the operator-sum representation. The Gaussian unitary in the dilation of an amplifier channel is a two-mode squeezing operator and its matrix elements were previously evaluated  to be \cite{kraus10}
\begin{align}
&T[{\cal A}(\kappa;n)]^{m_1,m_2}_{n_1,n_2} \nonumber\\
 &= {\kappa}^{-1} \sqrt{\frac{n_1!m_2!}{m_1!n_2!}} \sum_{r=0}^{n_2} \sum_{j=0}^{m_1} \, \binom{n_2}{r} \binom{m_1}{j}   (-1)^{r} (\kappa^{-1})^{n_2+j-r} \nonumber\\ &  ~~ \times (\sqrt{1-\kappa^{-2}})^{m_1+r-j} \, \,\delta_{n_1,r+j}\, \delta_{m_2,n_2+m_1-r-j},
\label{auni}
\end{align}
where $\binom{x}{y} = x!/(y! (x-y)!)$ is the binomial coefficient. As outlined in Sec. \ref{prelim} and Fig. \ref{fig4}, to compute the Kraus operators we set  $m_2 = \ell$, $n_2=n$, and eliminate $j$ using $\delta_{n_1,r+j}$  to obtain
\begin{align}
&T[{\cal A}(\kappa;n)]^{m_1,\ell}_{n_1,n} \nonumber\\ &=   {\kappa}^{-1} \sqrt{\frac{n_1!\ell!}{m_1!n!}}   \, g_1(n,n_1,m_1;\kappa) \,\delta_{\ell +n_1,m_1+n} ,
\end{align}
where
\begin{align}
&g_1(n,n_1,m_1;\kappa) \nonumber\\
&=  \sum_{r = \max \{0, n_1 -m_1\}}^{\min \{ n, n_1\}} \, \binom{n}{r} \binom{m_1}{n_1-r}  \, (-1)^{r} (\kappa^{-1})^{n+n_1-2r}\, \nonumber\\ &~~ \times (\sqrt{1-\kappa^{-2}})^{m_1+2r-n_1}.
\label{g2}
\end{align}
By  Eq. \eqref{kraus-rep}, we then obtain the Kraus operators  $A_{\ell}(\kappa;n)$ as 
\begin{align}
&A_{\ell}(\kappa;n)  =\kappa^{-1} \sum_{m_1,n_1 =0}^{\infty}  \sqrt{\frac{n_1!\ell!}{m_1!n!}} \, g_1(n,n_1,m_1;\kappa)  \nonumber\\ &~~~~~~~~~~~~\times\,\delta_{\ell,m_1+n-n_1}\, \ket{m_1} \bra{n_1} \nonumber\\
&= \kappa^{-1}\sum_{n_1 = \max\{ 0,n-\ell\} }^{\infty}  \sqrt{\frac{n_1!\ell!}{(\ell -n + n_1)!n!}}  \nonumber\\ &~~~~~ \times g_1(n,n_1,\ell-n+n_1;\kappa)  \, \ket{n_1 + (\ell - n)} \bra{n_1}.
\label{aosr}
\end{align}

\begin{figure}
\centering
\scalebox{0.45}{\includegraphics{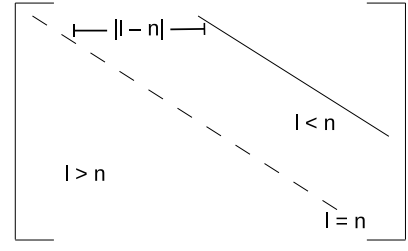}}
\caption{Structure of Kraus operators for the photon-added amplifier channels. Each Kraus operator has non-zero elements only along a line parallel to the diagonal whose location depends on $|\ell -n|$ where $\ell$ is the Kraus index and $n$ the number of photon addition. \label{fig5}}
\end{figure}

As a preliminary observation we see that each Kraus operator has non-zero elements only along a line parallel to the diagonal. We find a demarcation into three regions depending on $|\ell -n|$. For the case when the Kraus index $\ell < n$ this line falls above the diagonal and for $\ell > n$ it falls below the diagonal. For the case $\ell =n$ we see that the corresponding Kraus operator is diagonal in the Fock basis. This is schematically depicted in Fig. \ref{fig5}. If we set $n=0$ in Eq. \eqref{aosr} we obtain $\{A_{\ell}(\kappa;0)\}$ which are the Kraus operators for the quantum-limited amplifier channel ${\cal A}(\kappa;0)$~\cite{kraus10}.

\subsection{Photon-added attenuator ${\cal B}(\kappa;n)$}

\begin{figure}
\centering
\scalebox{0.75}{\includegraphics{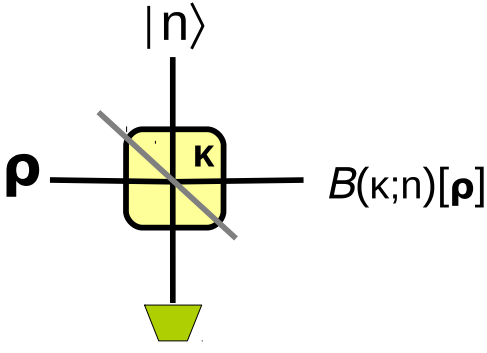}}
\caption{Realization of the photon-added attenuator channels ${\cal B}(\kappa;n)$. Here $\kappa$ denotes the channel parameter coming from the two-mode beamsplitter unitary and $n$ denotes the photon addition.\label{fig5b}}
\end{figure} 

\begin{figure}
\centering
\scalebox{0.45}{\includegraphics{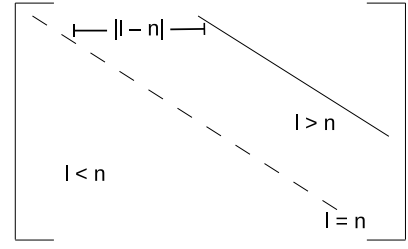}}
\caption{Structure of Kraus operators for the photon-added attenuator channels. Each Kraus operator with index $\ell$ has non-zero elements only along a line parallel to the diagonal, whose location depends on $|\ell -n|$, where $n$ is the photon-number state of the environment.\label{fig6}}
\end{figure}

We next consider the case of the photon-added attenuator channel which is depicted in Fig. \ref{fig5b}. The unitary in the dilation of the attenuator channel is the beamsplitter. The matrix elements of the two-mode beamsplitter unitary operator was previously evaluated as \cite{kraus10}
\begin{align}
&T[{\cal B}(\kappa;n)]^{m_1,m_2}_{n_1,n_2}  \nonumber\\ &
= \sqrt{\frac{m_1!m_2!}{n_1!n_2!}} \sum_{r=0}^{n_1} \sum_{j=0}^{n_2} \, \binom{n_1}{r} \binom{n_2}{j} \, (-1)^{n_2-j} \,\delta_{m_2,r+j} \nonumber\\ &~~~~\times \,(\kappa)^{n_1-r+j} \,(\sqrt{1-\kappa^2})^{r+n_2-j}  \delta_{m_1,n_1+n_2-r-j}\,.
\label{buni}
\end{align}
As in the earlier case, setting  $m_2 = \ell$,  $n_2=n$, and eliminating $j$ using $\delta_{m_2,r+j}$ we have
\begin{align}
T[{\cal B}(\kappa;n)]^{m_1,\ell}_{n_1,n} =  \sqrt{\frac{m_1!\ell!}{n_1!n!}} \,g_2(n,\ell,n_1;\kappa) \,\delta_{m_1,n_1+n-\ell} ,
\end{align}
where
\begin{align}
&g_2(n,\ell,n_1;\kappa)  =  \sum_{r = \max \{0, \ell -n\}}^{\min \{ \ell, n_1\}} \binom{n_1}{r} \,\binom{n}{\ell -r}\, (-1)^{n-\ell+r} \nonumber\\ & \hspace{2.2cm}\times (\kappa)^{n_1-2r+\ell}  \,  \,(\sqrt{1-\kappa^2})^{2r+n-\ell}.
\label{g1}
\end{align}
We then evaluate the Kraus operators  $B_{\ell}(\kappa;n)$ to be  
\begin{align} 
&B_{\ell}(\kappa;n)   = \sum_{m_1,n_1 =0}^{\infty}  \sqrt{\frac{m_1!\ell!}{n_1!n!}}   \,g_2(n,\ell,n_1;\kappa) \nonumber\\ &~~\hspace{2cm}~~\times \,\delta_{m_1,n_1+n-\ell}  \, \ket{m_1} \bra{n_1} \nonumber\\
&= \sum_{n_1 = \max \{0, \ell -n\}}^{\infty}  \sqrt{\frac{(n_1+n-\ell)!\ell!}{n_1!n!}} \,   g_2(n,\ell,n_1;\kappa)  \nonumber\\ & ~~~~~\, \times\ket{n_1 + (n-\ell)} \bra{n_1}.
\label{bosr}
\end{align}
We notice that as in the photon-added amplifier case each Kraus operator has non-zero elements only along a line parallel to the diagonal whose location depends on $|\ell -n|$ but the situation is now reversed when compared with the photon-added amplifier channel as depicted in Fig. \ref{fig6}.   This line is located in the upper triangle for $\ell >n$ and in the lower triangle for $\ell < n$.

\subsection{Photon-added phase conjugator ${\cal C}(\kappa;n)$}
\begin{figure}
\centering
\scalebox{0.75}{\includegraphics{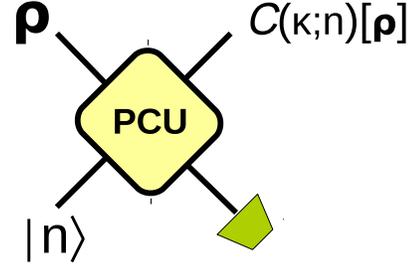}}
\caption{Realization of the photon-added amplifier channels ${\cal C}(\kappa;n)$. Here $\kappa$ denotes the channel parameter coming from the two-mode phase-conjugation unitary (PCU)  and $n$ denotes the photon addition.\label{fig6a}}
\end{figure}

\begin{figure}
\centering
\scalebox{0.45}{\includegraphics{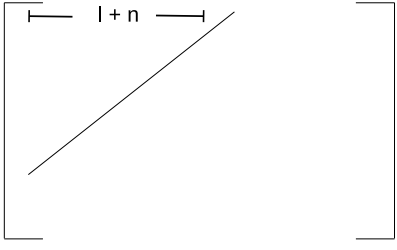}}
\caption{Structure of Kraus operators for the photon-added phase conjugation channels. Each Kraus operator has non-zero elements only along a particular `anti-diagonal' whose location depends on $\ell + n$, where $\ell$ is the Kraus index and $n$ is the value of photon addition. \label{fig7}}
\end{figure}

The last example we  consider is the photon-added phase-conjugation channel depicted in Fig. \ref{fig6a}. The unitary operator in the dilation of the phase-conjugation channel has matrix elements given by \cite{kraus10}
\begin{align}
&T[{\cal C}(\kappa;n)]^{m_1,m_2}_{n_1,n_2} \nonumber\\  
&= d_{\kappa}\sqrt{\frac{n_2!m_2!}{m_1!n_1!}} \sum_{r=0}^{m_1} \sum_{j=0}^{n_1} \, \binom{m_1}{r} \binom{n_1}{j} \, (\sqrt{1+\kappa^{2}})^{-(n_1+r-j)}  \,\, \nonumber\\
&~~\times (-1)^{j} (\sqrt{1+\kappa^{-2}})^{-(m_1-r+j)} \,\delta_{n_2,r+j} \delta_{m_2,n_1+m_1-r-j},
\label{cuni}
\end{align}
where $d_{\kappa}=(\sqrt{1+\kappa^{2}})^{-1}$ and the unitary is obtained by following the two-mode squeeze operator with a mode-flip operator, a fact that will be discussed in more detail in Sec.~\ref{phicomp}. As before by setting $m_2 = \ell$, $n_2=n$ and eliminating $j$ using $\delta_{n_2,r+j}$  we have
\begin{align}
&T[{\cal C}(\kappa;n)]^{m_1,\ell}_{n_1,n} =   d_{\kappa} \sqrt{\frac{n!\ell!}{m_1!n_1!}}   \,  g_3(n,n_1,m_1;\kappa) \,\delta_{\ell,m_1-n+n_1} ,
\end{align}
where
\begin{align}
&g_3(n,n_1,m_1;\kappa) \nonumber\\ &=  \sum_{r = \max \{0, n -n_1\}}^{\min \{ n, m_1\}} \, \binom{m_1}{r} \binom{n_1}{n-r} \,(\sqrt{1+\kappa^{-2}})^{-(m_1-2r+n)} \nonumber\\ &~~~~~\times \, (-1)^{n-r} \, (\sqrt{1+\kappa^{2}})^{-(n_1+2r-n)}.
\label{g3}
\end{align}
We then obtain the Kraus operators  $C_{\ell}(\kappa;n)$ as 
\begin{align}
C_{\ell}(\kappa;n)   &=  d_{\kappa}  \sum_{m_1,n_1 =0}^{\infty}  \sqrt{\frac{n!\ell!}{m_1!n_1!}}
\, g_3(n,n_1,m_1;\kappa) ~~~~~~~~~~~~~ \nonumber\\ & \hspace{3cm}\times \,\delta_{\ell,m_1-n+n_1}\, \ket{m_1} \bra{n_1} \nonumber\\
&= d_{\kappa}  \sum_{n_1 =  0 }^{n+\ell} \sqrt{\frac{n!\ell!}{(\ell + n-n_1)!n_1!}}  \,  \nonumber\\ &\times g_3(n,n_1,\ell+n-n_1;\kappa)    \, \ket{\ell + n - n_1} \bra{n_1}.
\label{e16}
\end{align}
We observe that $\{C_{\ell}(\kappa;n)\}$ are of finite rank for finite $n, \ell$. Furthermore, we note that unlike the photon-added amplifier and attenuation channels, each Kraus operator has non-zero elements along a line parallel to the `anti-diagonal' as depicted in Fig. \ref{fig7}.

We emphasize that for each of the photon-added amplifier, attenuator, and phase conjugation channels there exists an operator-sum representation in the Fock basis in which all the Kraus operators are real. We summarize the results of this section in the following remark.
\begin{theorem}
The collections $\{A_{\ell}(\kappa;n)\}$, $\{B_{\ell}(\kappa;n)\}$, $\{C_{\ell}(\kappa;n)\}$ given in Eqs. \eqref{aosr}, \eqref{bosr}, \eqref{e16} are a set of Kraus operators for the photon-added channels ${\cal A}(\kappa;n)$, ${\cal B}(\kappa;n)$, ${\cal C}(\kappa;n)$, respectively.  
\end{theorem}

\section{Properties of photon-added channels}
In this section we derive various properties of photon-added channels whose operator-sum representation we derived in the previous section. We  emphasize the fact that the Fock basis plays an important role to describe the properties of these non-Gaussian channels. We now begin with the action of the photon-added channels in the Fock basis.

\subsection{Action on Fock states \label{fockact}}
From the operator-sum representation of each of the photon-added channels we study their action on the Fock basis and this turns out to reveal many important structural properties. We first begin with the action of the photon-added attenuator channel on a given Fock state $\proj{j}$. We have from  Eq. \eqref{bosr} 
\begin{align}
& {\cal B}(\kappa;n)[\proj{j}] =\sum_{\ell=0}^{\infty} B_{\ell}(\kappa;n)  \proj{j} B_{\ell}(\kappa;n)^{\dagger} \nonumber\\
&= \sum_{\ell=0}^{\infty} \,\sum_{n_1 = \max\{0,n-\ell\}}^{\infty} \,\sum_{m_1 = \max\{0,n-\ell\}}^{\infty}  
 \sqrt{\frac{(n_1+n-\ell)!\ell!}{n_1!n!}}\,    \nonumber\\
&~~\times \sqrt{\frac{(m_1+n-\ell)!\ell!}{m_1!n!}} g_2(n,\ell,n_1;\kappa)\,  g_2(n,\ell,m_1;\kappa) \braket{n_1}{j}  \nonumber\\ & ~~~~~~~\times \braket{j}{m_1} \ket{n_1 + n-\ell}\bra{m_1 + n-\ell}   \nonumber\\
&= \sum_{\ell=0}^{j+n} \frac{(j+n-\ell)!\ell!}{j!n!} g_2(n,\ell,j;\kappa)^{2} \proj{j+n-\ell}.
\label{bfock} 
\end{align}
We observe that states diagonal in the Fock basis are mapped to states that are also diagonal in the Fock basis irrespective of the amount of photon addition. Furthermore, we see that states having a finite Fock support are also taken to states with finite Fock support. An analogous observation for the Gaussian channel case $n=0$~\cite{kraus10} played an important role in the resolution of an  output entropy optimization problem\cite{gio17}.

We next consider the case of the photon-added amplifier and consider its action on an input Fock state $\proj{j}$. We have by  Eq.~\eqref{aosr} 
\begin{align}
&{\cal A}(\kappa;n)[\proj{j}] =  \sum_{\ell=0}^{\infty} A_{\ell}(\kappa;n) \proj{j} A_{\ell}(\kappa;n)^{\dagger} \nonumber \\
&= \kappa^{-2}  \sum_{\ell=0}^{\infty} \sum_{n_1 = n_0 }^{\infty} \sum_{m_1 = n_0 }^{\infty}\sqrt{\frac{n_1!\ell!}{(\ell -n + n_1)!n!}}\,\,  \braket{n_1}{j} \, \braket{j}{m_1 }\nonumber\\ 
&~~\times \,\sqrt{\frac{m_1!\ell!}{(\ell -n + m_1)!n!}}   ~g_1(n,n_1,\ell-n+n_1;\kappa) \nonumber\\ 
& \times \,g_1(n,m_1,\ell-n+m_1;\kappa) \, \ket{n_1 + \ell - n} \bra{m_1+\ell - n}, \nonumber\\ &~~~~~~~~~~~~n_0= \max\{ 0,n-\ell\}.
\end{align}
The above expression evaluates to 
\begin{align}
&{\cal A}(\kappa;n)[\proj{j}] \nonumber\\ &=  \kappa^{-2}  \sum_{\ell= \max\{n-j,0\}}^{\infty} \frac{j!\ell!}{(\ell -n + j)!n!} \nonumber\\
&\,\times  g_1(n,j,\ell-n+j;\kappa)^2  \,\proj{\ell -n + j}.
\label{afock}
\end{align}

\begin{table}
\caption{Largest and smallest Fock state support of an output state $\hr{\rm out}$ for a given input state $\hr{\rm in}$ for each of the photon-added amplifier, attenuator, and phase-conjugation channels. Here $N_{\rm max}[\hr{\rm }]$ and $N_{\rm min}[\hr{\rm }]$ denote the largest and smallest Fock state in the support of  $\hr{\rm }$.}
\begin{tabular}{lcc}
\hline
{\bf Channel} & $\bs{N_{\rm min}[\hr{\rm out}]}$ & $\bs{N_{\rm max}[\hr{\rm out}]}$  \\
\hline
\hline 
${\cal A}(\kappa;n)$  &~~~$\max \{ N_{\rm min}[\hr{\rm in}] -n,0\}$ & $\infty$\\
${\cal B}(\kappa;n)$ &$0 $& ~~~$N_{\rm max}[\hr{\rm in}] + n$\\
${\cal C}(\kappa;n)$ & ~~~$\max \{ n-N_{\rm min}[\hr{\rm in}],0\}$ &$\infty$\\
\hline
\end{tabular}
\label{table1}
\end{table}

Finally, for the case of the action of the photon-added phase-conjugation channel on an input Fock state,  we have by  Eq. \eqref{e16} 
\begin{align}
&{\cal C}(\kappa;n)[\proj{j}] =  \sum_{\ell=0}^{\infty} C_{\ell}(\kappa;n) \proj{j} C_{\ell}(\kappa;n)^{\dagger} \nonumber \\
&= d^2_{\kappa}  \sum_{\ell =0}^{\infty} \sum_{m_1 =  0 }^{n+\ell} \sum_{n_1 =  0 }^{n+\ell} \sqrt{\frac{n!\ell!}{(\ell + n-n_1)!n_1!}}  \, \braket{n_1}{j} \braket{j}{m_1}   \nonumber\\ 
&~ \times \,\, \sqrt{\frac{n!\ell!}{(\ell + n-m_1)!m_1!}} \,  g_3(n,n_1,\ell+n-n_1;\kappa) \nonumber\\ &\times \,g_3(n,m_1,\ell+n-m_1;\kappa)  \, \ket{\ell + n - n_1} \bra{\ell + n - m_1}\nonumber\\
&=d^2_{\kappa} \sum_{\ell= \max\{j-n,0\}}^{\infty} \frac{n!\ell!}{(\ell + n-j)!j!} \, g_3(n,j,\ell+n-j;\kappa)^2 \nonumber\\ &~~~~~~~~ \times  \proj{\ell +n - j}.
\label{cfock}
\end{align} 
From the action on an arbitrary Fock state one can deduce some properties of the action of the photon-added channels on a general state $\hr{}$. For this let us denote by $N_{\rm max}[\hr{}]$ and $N_{\rm min}[\hr{}]$ the largest and smallest Fock state in the support of some given state $\hr{\rm }$. 

Let an input state $\hr{\rm in}$ to either of the photon-added channels be mapped to an output state $\hr{\rm out}$. Then by  Eqs. \eqref{bfock}, \eqref{afock}, and \eqref{cfock} we have the following proposition regarding the action of photon-added channels on $\hr{\rm in}$ and this is also summarized in Table \ref{table1}.   
\begin{proposition}
The $n$ photon-added attenuator produces an output with $N_{\rm max}[\hr{{\rm out}}] = N_{\rm max}[\hr{\rm in}] + n$ and  $N_{\rm min}[\hr{{\rm out}}] =0$,  the $n$ photon-added amplifier produces an output with $N_{\rm min}[\hr{{\rm out}}] = \max \{ N_{\rm min}[\hr{\rm in}] -n,0 \}$ and  $N_{\rm max}[\hr{{\rm out}}] = \infty $, and finally, the $n$ photon-added phase conjugator outputs a state with  $N_{\rm min}[\hr{{\rm out}}] = \max \{ n-N_{\rm min}[\hr{\rm in}],0 \}$ and $N_{\rm max}[\hr{{\rm out}}] = \infty$.
\label{ranges}
\end{proposition}
So we see the effect of photon addition in respect of the restrictions on the Fock-support ranges for the outputs of each of the channels. Also we note that photon addition gives rise to different properties of the output states as compared with the Gaussian ($n=0$) case through its dependence on $n$. As a consequence of Proposition \ref{ranges} we observe the following.
\begin{corollary}
There does not exist any fixed point with finite support in the Fock basis for the photon-added attenuator, amplifier and phase conjugation channels.
\end{corollary}
For a comparison with the fixed points of the corresponding Gaussian amplifier, attenuator, and phase-conjugation channels, i.e., in the absence of any photon addition ($n=0$), we refer the reader to \cite{kraus10}. As a final comment, the action of the channels on input Fock states helps us understand aspects of Fock-basis coherence which we discuss in Section \ref{fcoh}.

\subsection{Activation of quantum-optical nonclassicality}
We now present a consequence of photon addition in respect of  nonclassicality in the quantum-optical context and it is well-known that nonclassicality of light is an important resource in many situations like entanglement generation\cite{ntoE1,ntoE2,ntoE3} and use in laser interferometers\cite{roman}. We illustrate with a useful example that nonclassicality can be activated at the outputs of certain channels, through our scheme of photon addition to the environment state, that would otherwise output only classical states. Before we begin with a specific example we  briefly recall the notion of quantum-optical nonclassicality that we are dealing with in this section. 

Any state $\hr{}$ can be  described in terms of its Glauber-Sudarshan diagonal `weight' function $\phi(\bs{\alpha};\hr{})$~\cite{ecg}:
\begin{align}
\hr{} = \int\frac{d^{2n} \bs{\alpha}}{\pi^n}\, {\phi} (\bs{\alpha};\hr{}) | \bs{\alpha} \rangle \langle \bs{\alpha} |.
\label{nonc}
\end{align} 
Then $\hr{}$ is said to be classical if its associated $\phi(\bs{\alpha};\hr{})$ function is everywhere non-negative over the phase-space $\bs{\alpha} \in \mathbb{R}^{2n} \simeq \mathbb{C}^{n}$, else it is said to be nonclassical. In other words, a  classical state is  a convex mixture of coherent states.

It is known that the quantum-limited phase conjugation channel is nonclassicality breaking \cite{kraus10,nbm,nb,pon}. In other words, the output corresponding to any input to the phase-conjugation channel is rendered classical. Let us now consider the effect of photon addition, i.e. the corresponding photon-added phase conjugator, on the output states of the corresponding channel and we have the following proposition. 
\begin{proposition}
The photon-added phase conjugation channel ${\cal C}(\kappa;n)$ is  nonclassicality breaking only for $n=0$. For $n>0$, i.e. under any non-trivial photon addition,  nonclassicality is activated at the output of the corresponding channel. 
\label{nogothm}
\end{proposition}
\proof As mentioned earlier it is known that the phase conjugation channel ${\cal C}(\kappa;0)$ (with no photon addition) is nonclassicality breaking \cite{kraus10,nb,nbm,pon}. We now consider the case $n \neq 0$. Let us consider as input the vacuum state $\proj{0}$. Having seen the action of the photon-added channel in the Fock basis, we have by  Eq. \eqref{cfock} that the corresponding output is given by 
\begin{align}
&{\cal C}(\kappa;n)[\proj{0}] \nonumber\\&= d^2_{\kappa} \sum_{\ell= 0}^{\infty} \frac{n!\ell!}{(\ell + n)!}\, g_3(n,0,\ell+n;\kappa)^2 \proj{\ell +n}.
\end{align}  
So it clear that 
\begin{align}
\bra{\tau} {\cal C}(\kappa;n)[\proj{0}]\ket{\tau} = 0 ~ \forall~ \tau = 0,1,\cdots, n-1,  
\end{align}
and in particular for $\tau =0$. This implies further that the $Q$-function defined as 
$Q(\bs{\alpha};\hr{}) = \bra{\bs{\alpha}}\hr{} \ket{\bs{\alpha}}$  is zero when evaluated at $\bs{\alpha}=0$. It then immediatly follows that the corresponding output state 
${\cal C}(\kappa;n)[\proj{0}]$ is nonclassical for each $n > 0$ \cite{lut}. Hence we see that photon addition to the environment has activated nonclassicality at the output of the corresponding photon-added phase-conjugation channel. $\blacksquare$

We have shown that nonclassicality can be activated at the output of a phase-conjugation channel through photon addition in the ancilla port vacuum state.  We also highlight that this activation is measure-independent since it deals directly with the notion of quantum-optical nonclassicality described in Eq. \eqref{nonc}. We now extend this property to all input classical states thereby proving a form of strong-activation. 
\begin{theorem}
\label{actcl}
All classical states are mapped to nonclassical states under the action of the photon-added phase conjugation channel ${\cal C}(\kappa;n)$ for every $n > 0$.
\end{theorem}
\proof The crucial observation is that for any classical state $\hr{\rm in}$, i.e., a state with everywhere non-negative diagonal function $\phi(\bs{\alpha};\hr{\rm in})$, the support on the vacuum state is strictly positive. In other words $\bra{0} \hr{} \ket{0} = Q(0) > 0$. If this is not the case, then the state is nonclassical as mentioned in the proof of Proposition \ref{nogothm}.

Therefore for any classical state the minimum Fock state with non-zero Fock support is zero, i.e., $N_{\rm min}[\hr{\rm in}] =0$. Then by Proposition \ref{ranges} or equivalently Table \ref{table1} we have that \begin{align} N_{\rm min}[{\cal C}(\kappa;n)(\hr{\rm in})] = n - N_{\rm min}[\hr{\rm in}] = n. \end{align} So  $\bra{0} {\cal C}(\kappa;n)(\hr{\rm in}) \ket{0}=0$ for $n \geq 1$. By arguments used in the proof of Proposition \ref{nogothm} we have that the output corresponding to a classical input for the photon-added phase conjugation channel is nonclassical. \hfill $\blacksquare$ 


\subsection{Complementarity \label{phicomp}}
Consider a channel $\Phi$ which has a Stinespring representation as given in  Eq. \eqref{e1}. Then the complementary channel, denoted by $\Phi^{c}$, is given by tracing out the system degrees of freedom, i.e., 
\begin{align}
\Phi^c[\hr{A}] = {\rm Tr}_A\, \left[ U_{AE}\,(\hr{A} \otimes \ket{\psi}_E \bra{\psi})\,U_{AE}^{\dagger} \right].
\end{align} 
Therefore the complementary channel maps states of the system to states of the environment. We make the following observation with respect to complementarity of photon-added channels. 

\begin{figure}
\centering
\includegraphics[width=\columnwidth]{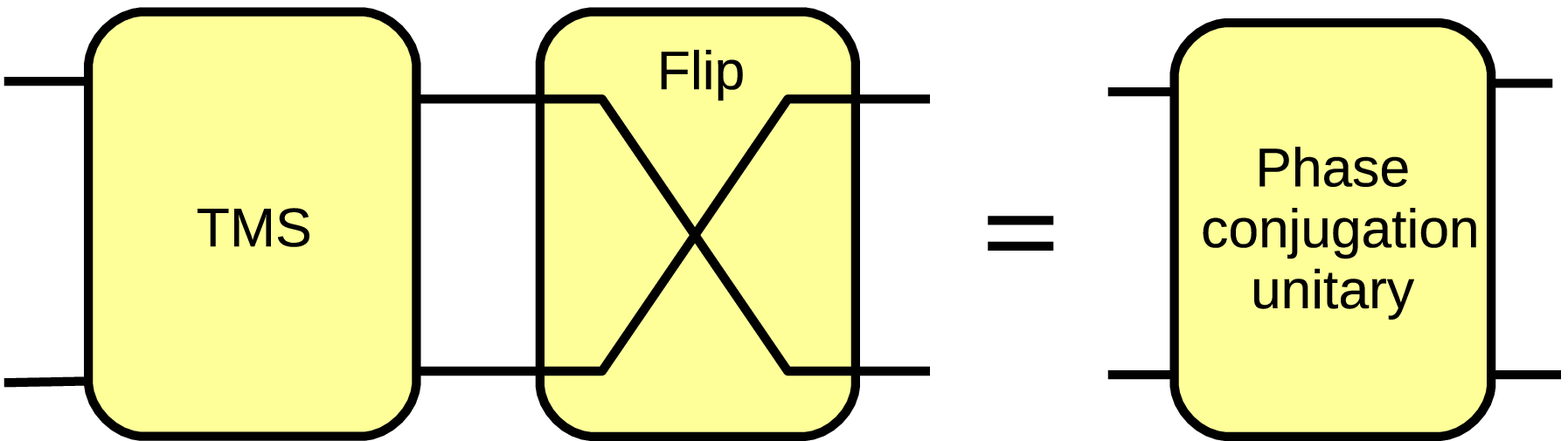} \vspace{0.5cm} \includegraphics[width=\columnwidth]{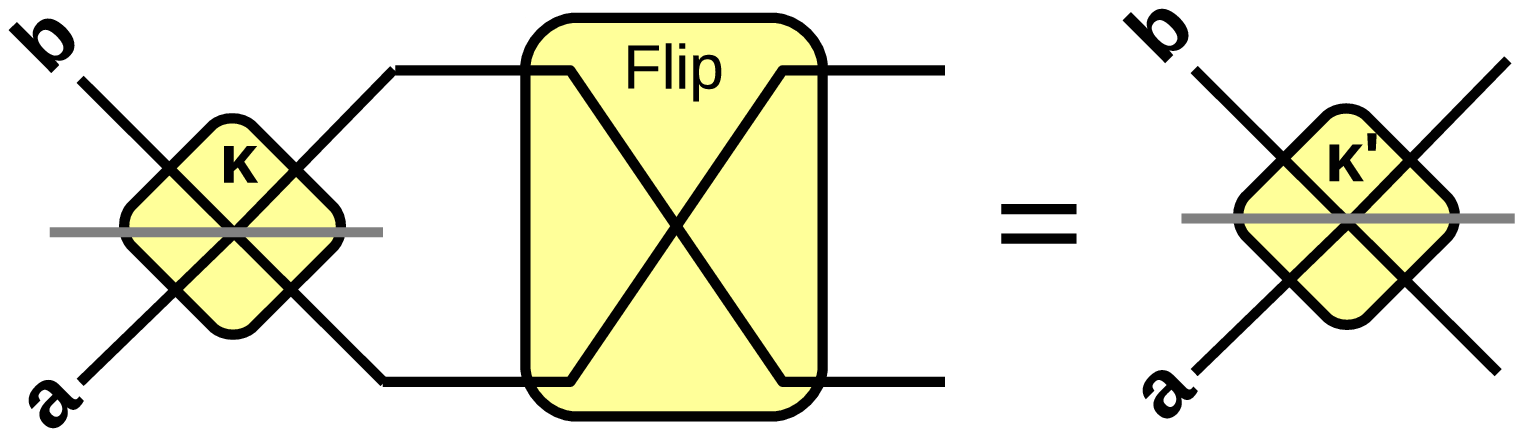}
\caption{Relation between the unitary operators in the Stinespring dilation of the amplifier channel denoted by TMS (two-mode squeeze operator) and that of the phase conjugation channel with parameters $\kappa$ and $\kappa^{\prime} = \sqrt{\kappa^{2}-1}$ (above), and the  two unitaries (beamsplitters) in the Stinespring dilation of the corresponding attenuator channels with parameters $\kappa$ and $\kappa^{\prime} =\sqrt{1-\kappa^2}$ respectively (below). The pair-wise unitaries in each case are connected by the mode-flip operator. 
\label{fig9}}
\end{figure}

\begin{proposition}
The complementary channel corresponding to the photon-added amplifier channel ${\cal A}(\kappa;n)$ with $\kappa \geq 1$ is the photon-added phase-conjugation channel ${\cal C}(\sqrt{\kappa^{2}-1};n)$ for every $n$. Similarly, the photon-added attenuator ${\cal B}(\kappa;n)$ with $0 \leq \kappa \leq 1$  is complementary to another photon-added attenuator ${\cal B}(\sqrt{1-\kappa^2};n)$ for every $n \geq 0$. 
\label{compcha}
\end{proposition}
\proof The proof follows by comparing the matrix elements of the corresponding unitary operators in the dilations as explained in Appendix \ref{compli} and schematically depicted in Fig.~\ref{fig9}.
\hfill $\blacksquare$ 

We have a straightforward implication regarding the minimum output entropies.
\begin{corollary}
The minimum output entropy of the photon-added amplifier channel ${\cal A}(\kappa;n)$ and the photon-added phase-conjugation channel ${\cal C}(\sqrt{\kappa^{2}-1};n)$ are identical. Similarly the minimum output entropy of beamsplitters ${\cal B}(\kappa;n)$ and ${\cal B}(\sqrt{1-\kappa^2};n)$ are identical. 
\end{corollary}
\noindent This can be seen easily from two facts. First, the minimum output entropy is achieved on input states that are  pure. Second, for pure inputs the channel and its complementary have identical output entropy, and therefore also holds true for the pure optimal state that minimizes the output entropy of the channel.
\hfill $\blacksquare$  

\subsection{Activation of quantum and private capacities}
We now make a few remarks regarding the quantum and private classical  capacities of the photon-added phase-conjugation channels and present a novel phenomenon of activation of these capacities. Given a quantum channel one of its defining characteristics is its usefulness in transmitting information. Depending on the resources available and the specific information-theoretic task to accomplish, a host of capacities have been discussed and studied in literature. We  focus on the fundamentally important quantum and private classical capacities \cite{wildebook}.

The quantum capacity ${\mathcal Q}(\Phi)$ of a channel $\Phi$ is the rate (qubits per channel use) at which the channel can reliably transmit quantum information in the asymptotic limit of channel uses, and can be expressed as 
\begin{align}
{\cal Q}(\Phi) &= \lim_{n \to \infty} \frac{1}{n}{\cal Q}^{(1)}(\Phi^{\otimes n}), \nonumber \\  
{\cal Q}^{(1)}(\Phi) &= \sup_{\hr{}} I_{\rm coh}(\Phi,\hr{}),  \nonumber  \\ 
I_{\rm coh}(\Phi,\hr{}) &= S(\Phi(\hr{})) - S(\Phi^c(\hr{})), 
\label{quandef}
\end{align}
where $I_{\rm coh}(\Phi,\hr{})$ denotes the coherent information for a given channel $\Phi$ and input state $\hr{}$ \cite{schumi96}, and $S(\cdot)$ stands for the von Neumann entropy of the state and $\Phi^{c}$ stands for the complimentary channel (see Section \ref{phicomp}).

On the other hand, the private capacity ${\cal P}(\Phi)$ is the rate (bits per channel use) at which $\Phi$ can transmit classical information reliably in the asymptotic number of channel uses such that the information is secret from an eavesdropper, and can be expressed as 
\begin{align}
{\cal P}(\Phi) &= \lim_{n \to \infty} \frac{1}{n} {\cal P}^{(1)}(\Phi^{\otimes n}),  \nonumber \\
{\cal P}^{(1)}(\Phi) &= \sup_{p_x,\,\hr{x}} I([1\!\!1 \otimes \Phi](\hat{\sigma})) - I([1\!\!1 \otimes \Phi^{c}](\hat{\sigma})),  
\label{pridef}
\end{align} 
where $\hat{\sigma} = \sum_x p_x \ket{x}\bra{x} \otimes \hr{x}$ is a classical-quantum (cq) state and $I(\hr{AB}) = S(\hr{A}) + S(\hr{B}) - S(\hr{AB})$ is the mutual information of a bipartite state $\hr{AB}$. Further, it is also known that both the quantum and private classical capacity are zero for entanglement breaking channels \cite{shor04,wildebook}.  

In respect of the phase-conjugation channel ${\cal C}(\kappa;0)$ we know that it is not only nonclassicality breaking but also entanglement breaking \cite{holevo08,nb}. This implies further that the quantum capacity and the private classical capacity of ${\cal C}(\kappa;0)$ is zero \cite{nbm}. We have seen by Theorem \ref{actcl} that the corresponding photon-added channels ${\cal C}(\kappa;n)$ are not nonclassicality breaking for all $n>0$. Hence we take this as evidence that not only nonclassicality is activated at the output of the phase-conjugation channel after photon addition applied on the ancilla state but also that both the quantum and private classical capacities are activated as well. We find that indeed this is the case with specific numerical examples. 

We  have seen in Sec. \ref{phicomp} that for the photon-added phase conjugation channel ${\cal C}(\kappa;n)$, the complementary channel is given by the photon-added amplifier channel ${\cal A}(\sqrt{\kappa^2+1};n)$. This is important for evaluating the coherent information $I_{\rm coh}(\Phi,\hr{})$ and appears in the evaluation of the quantum capacity of $\Phi$ as given in Eq. \eqref{quandef}, where $\Phi^c$ denotes the complementary channel of $\Phi$.

\subsubsection{Numerical examples}
\begin{figure}
\centering
\includegraphics[width=\columnwidth]{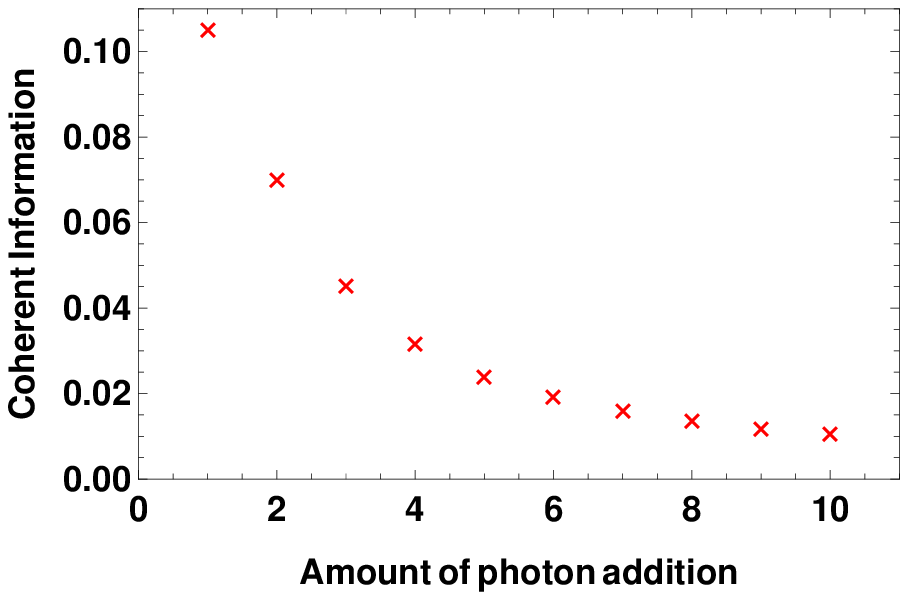}
\caption{Coherent information $I_{\rm coh}({\cal C}(\kappa;n),\hr{})$ as a function of photon addition $n$ with $\kappa = \sqrt{1.5^2-1}$ and $\hr{} = 0.6 \proj{0} + 0.4 \proj{1}$. \label{cohinfo}}
\end{figure}

\begin{figure}
\centering
\includegraphics[width=\columnwidth]{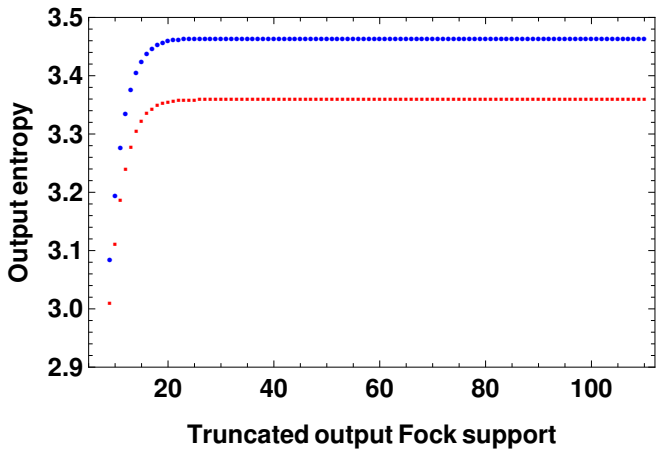}
\caption{Variation of entropy as a function of the truncated value of the output Fock state. The top  line (in blue with circle plot points) corresponds to $S({\cal C}(\kappa;1)[\hr{}])$, $\kappa = \sqrt{1.5^2 -1}$ and the bottom line (in red with square plot points) corresponds $S({\cal A}(1.5;1)[\hr{}])$, with $\hr{} = 0.6 \proj{0} + 0.4\proj{1}$. \label{trunc}}
\end{figure}

As our first example we consider the photon-added phase conjugation channel ${\cal C}(\kappa;n)$, $\kappa = \sqrt{1.5^2-1} = 1.11803$ and $n$ varying,  with input state  $0.6 \proj{0} + 0.4 \proj{1}$ (diagonal in the Fock basis). The corresponding photon-added amplifier channel has $\kappa^{\prime}=1.5$ with the same value of $n$. We compute the coherent information for this state and we depict the same in Fig.~\ref{cohinfo} for $n$ varying from $1 \to 10$, values that are of practical relevance as well.
We see that for this input state the coherent information decreases with increasing value of photon addition. Further for $n=0$ or no photon addition, the coherent information evaluates to $-0.2239$, as expected. We note that the coherent information provides a lower bound to the quantum capacity \eqref{quandef}, i.e.,
\begin{align}
{\cal Q}(\Phi) \geq I_{\rm coh}(\Phi,\hr{}),
\end{align}
and hence the quantum capacity $Q(({\cal C}(\kappa;n))>0$ for $\kappa = \sqrt{1.5^2-1}$ and $n = 1,2,\cdots,10$. We  note that the output states of the photon added phase conjugation channels are of infinite rank and hence their numerical evaluation required a truncation of the output Hilbert space. For the case $n = 1$, Fig. \ref{trunc} depicts the entropy $S({\cal C}(\kappa;1)[\hr{}])$, $\kappa = \sqrt{1.5² -1}$ and $S({\cal A}(1.5;1)[\hr{}])$ with $\rho = 0.6 \proj{0} + 0.4\proj{1}$ as a function of the truncated value of the output Fock state. The numerically evaluated entropies in Figures \ref{cohinfo} and Table \ref{table2} (that we explain in the next paragraph) were for output states truncated at value $n = 110$. 
\begin{table*}
\caption{Output entropies and coherent information for various input states and values of photon addition. Here $\kappa = \sqrt{1.5²-1}$ and $\kappa^{\prime} = 1.5$. Note that ${\cal A}(\kappa^{\prime};n)$ is the complementary channel to ${\cal C}(\kappa;n)$. Note that $I_{\rm coh}(\Phi) >0 \implies {\cal Q}(\Phi) >0$. }
\begin{tabular}{c|c|c|c|c}
\hline
~~$n$~~ & ~~$\hr{\rm in}$~~ & ~~$S({\cal C}(\kappa;n)[\hr{\rm in}])$~~ &~~$S({\cal A}(\kappa^{\prime};n)[\hr{\rm in}])$~~ & ~~$I_{\rm coh}({\cal C}(\kappa;n),\hr{\rm in})$   \\
\hline
\hline 
1 & $0.6 \proj{0} + 0.4 \proj{1}$&3.46527 &3.35954 &0.10573 \\ 
2 & $0.4 \proj{0} + 0.3 \proj{1} + 0.4 \proj{2}$ & 4.12843& 4.02385& 0.10458 \\
3 & $0.35 \proj{0} + 0.35 \proj{1} + 0.3 \proj{2}$ & 4.36985&4.27853 & 0.09132\\
4 & $0.35 \proj{0} + 0.35 \proj{1} + 0.3 \proj{2}$ & 4.54768&4.48032 & 0.06736\\
5 & $0.3 (\proj{0} + \proj{1}) + 0.2(\proj{2} + \proj{3})$ &4.86302 & 4.78464& 0.07838\\
\hline
\end{tabular}
\label{table2}
\end{table*}

As a second example we consider different input states that are all diagonal in the Fock basis and evaluate the corresponding entropies similar to the method outlined above and tabulated in Table \ref{table2}. We  reiterate that any positive coherent information $I_{\rm coh} > 0$ is a already a lower bound to the quantum capacity. 

Hence we see that photon addition to the environment (vacuum) state leads to the activation of quantum capacity. Since the quantum capacity is a lower bound to the private capacity~\cite{wildebook}, i.e. \begin{align}{\cal Q}(\Phi) \leq {\cal P}(\Phi),\end{align} we have that the private capacity is also activated for the phase conjugation channel. This is a clear phenomenon where no amount of photon addition (or any operation) on the input state to the phase conjugation channel would give a non zero quantum capacity whereas even a single photon addition to the environment already activates its quantum and private capacities. 

\begin{figure}
\centering
\includegraphics[width=\columnwidth]{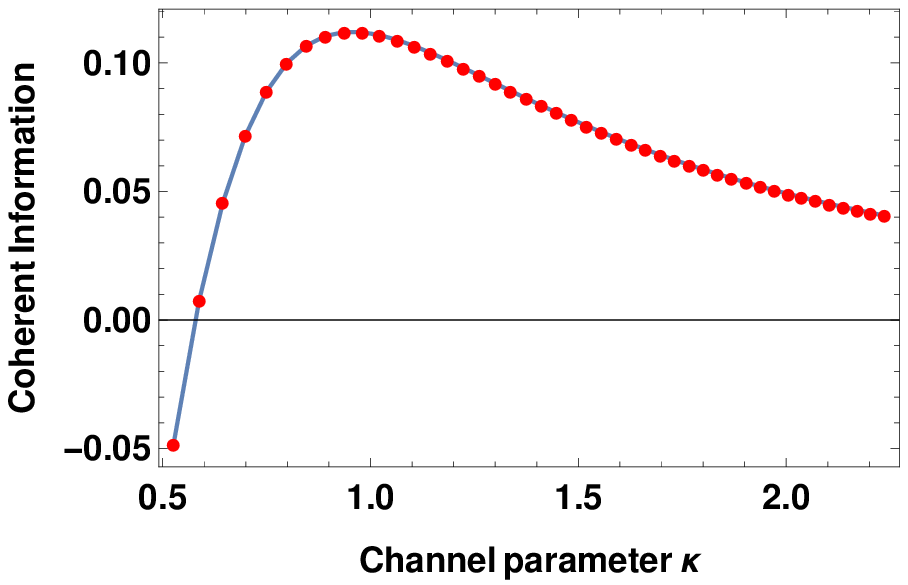}
\caption{Variation of coherent information $I_{\rm coh}({\cal C}(\kappa;1),\hr{\rm in}) $ as a function of the channel parameter $\kappa$ with $\hr{\rm in} = 0.6 \proj{0} + 0.4\proj{1}$ and $n=1$. \label{varyk}}
\end{figure}

As a final example we fix both the input state $\hr{\rm in}= 0.6 \proj{0} + 0.4 \proj{1}$ and the amount of photon addition to $n =1$. But now we vary the channel parameter $\kappa$ and we plot the coherent information in Fig. \ref{varyk}. 
We  note that our examples were constructive in nature and we believe that this phenomenon holds true for all values of $\kappa >0$ and $n$ for the photon added phase conjugation channel ${\cal C}(\kappa;n)$ and we state this as the following conjecture. 
\begin{conjecture}
The quantum capacity ${\cal Q}({\cal C}(\kappa;n))$ and the private classical capacity ${\cal P}({\cal C}(\kappa;n))$ 
for the photon-added phase-conjugation channels ${\cal C}(\kappa;n)$  is positive for all $n>0$. 
\end{conjecture}
If this conjecture is proven it would imply that the quantum and private classical capacities of the phase-conjugation channels are activated alongside the output nonclassicality.  This could have far-reaching practical implications for quantum optical communication.

\subsection{Fock-basis coherence and resource theories\label{fcoh}}
 The study of the resource theory of coherence has garnered much attention in recent literature \cite{dong,plenio,eric} with motivation taken from the resource theory of entanglement. A crucial element in these resource theories is the description of the set of free operations, i.e., operations that can be implemented without incurring costs, for manipulation among resource states, LOCC (local operations and classical communication) being an example in the resource theory of entanglement. We make a few remarks regarding the resource theory of coherence in the Fock basis and photon-added Gaussian channels.  This is further application of the action of the channel in the Fock basis considered in Section \ref{fockact}. 

Let us denote the set of operators diagonal in the Fock basis  by ${\cal D}$, i.e.,
\begin{align}
{\cal D} := \left \{ \hat{\sigma} = \sum_{n=0}^{\infty} p_n \ket{n} \bra{n}, \, p_n \geq 0\,\forall\, n\right\}.
\end{align} From the action of the photon-added channels in the Fock basis we can obtain its behaviour with respect to generating coherence in the Fock basis. By  Eqs. \eqref{bfock}, \eqref{afock}, and \eqref{cfock} we immediately see that states diagonal in the Fock basis are taken to states that are diagonal in the Fock basis. Such channels are known as Fock-preserving channels \cite{raul16} and we then have the following proposition. 
\begin{proposition} The photon-added attenuator, amplifier and phase conjugation channels are Fock-preserving, i.e. $\Phi(\hr{}) \in {\cal D}$ for $\hr{}\in {\cal D}$.
\label{free}
\end{proposition}
In the language of resource theory of coherence \cite{dong,plenio,eric}, states that are diagonal in the Fock basis (the basis under consideration) are known as incoherent states (which are the common free states). 
Operations that map incoherent states to incoherent states are known as maximally incoherent operations. By Proposition \ref{free} we have that the photon-added amplifier, attenuator, and phase-conjugation channels are all maximally incoherent operations for every $n\geq 0$. 

We can refine the specific class to which these non-Gaussian operations belong. From the structure of the Kraus operators, we have by  Eqs. \eqref{aosr}, \eqref{bosr}, and \eqref{e16} that each individual Kraus operator takes states diagonal in the Fock basis to operators diagonal in the Fock basis, i.e. $F_{\ell}\,\hr{}\, F_{\ell}^{\dagger} \in {\cal D}$ for $\hr{}\in {\cal D}$. 
Operations where not only the map as a whole but also each Kraus operator takes incoherent states to incoherent states are known as incoherent operations. Thus the photon-added channels we consider are not only maximally incoherent but also incoherent operations \cite{eric}. 

Furthermore, it is also easy to see that the action of each individual Kraus operator commutes with the completely dephasing map $\triangle$ (in the Fock basis) which is defined as 
\begin{align}
\triangle(\hr{}) = \sum_{n=0}^{\infty} \bra{n} \hr{} \ket{n} \proj{n},
\end{align}
i.e., $\triangle (F_{\ell}\,\hr{}\,F_{\ell}^{\dagger}) = F_{\ell}\left( \triangle (\hr{}) \right)F_{\ell}^{\dagger} \ \forall \ell$, for $F = A$, $B$, $C$. Channels for which each individual Kraus operator action commutes with the completely dephasing map are known as strictly incoherent operations. Due to this fact the photon-added channels that we have considered belong to this class, and we state it in the following remark. 
\begin{remark}
The photon-added attenuator, amplifier and phase conjugation channels belong to the class of strictly incoherent maps acting on the Fock basis for each $n\geq 0$. 
\end{remark}
Note that the set of strictly incoherent operations is contained in the set of incoherent operations, which is in turn contained in the set of maximally incoherent operations \cite{eric}. 

We now observe a few properties of the photon-added channels with respect to other resources. We have seen that the photon-added channels are Fock-preserving or coherence non-generating in the Fock basis. Further, when the resource is quantum-optical nonclassicality, by Theorem \ref{actcl} the photon-added phase-conjugation channel is resource generating, since its action on classical input states generates nonclassicality at the output. As a third example of a resource let us consider non-Gaussianity. Then we see by Eq. \eqref{e4} that any  Gaussian input state is necessarily taken to a non-Gaussian state at the output. In other words, the photon-added phase conjugation, amplifier, and attenuator channels are all resource generating where input Gaussian states are taken to non-Gaussian states. To summarize the photon-added channel are resource non-generating on incoherent Fock states and resource generating on Gaussian states. Further, examples of channels exist which are resource generating on classical states.

\subsection{Environment-assisted error correction \label{envasec}}
\begin{figure}
\centering
\includegraphics[width=\columnwidth]{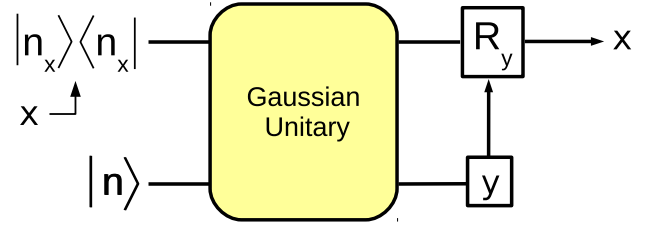}
\caption{Scheme for environment-assisted error correction for transmitting classical information through the photon-added attenuator, amplifier, and phase conjugation channels. Here the label $x$ stands for classical information encoded into Fock states and input to the channel. Conditioned on classical feedback $y$ from the environment that results from a measurement, a recovery map $R_y$ is applied at the output of the channel to achieve error correction. \label{fig8}}
\end{figure}

We now briefly discuss the notion of environment-assisted error correction for photon-added Gaussian channels. We follow the formalism that was introduced by Gregoratti and Werner in Ref. \cite{grego1}. A schematic of the protocol is provided in Fig. \ref{fig8}. Classical information is encoded in some chosen basis states and sent into the communication channel. We assume that there is classical feedback from the environment which contains information  about the exact Kraus operator that has acted on the input state based on a measurement outcome. Conditioned on this information, one can apply a recovery map at the output of the channel to correct for loss of information. 

It turns out that the structure of the quadratic operators $\{W_{\ell \ell} = F_{\ell}^{\dagger} F_{\ell}\}$ detailed in the Appendix Sec. \ref{quad}, where $\{F_{\ell}\}$ are the Kraus operators of a given channel, has consequences for the existence of an error correction scheme for environment assisted information transmission. We first begin with a proposition regarding these quadratic operators corresponding to photon-added Gaussian channels that follows from Eqs. \eqref{wc}, \eqref{wb}, and \eqref{wa} in the Appendix \ref{quad}. 
\begin{proposition}
The positive operators $\{W[\Phi]_{\ell \ell} = F_{\ell}^{\dagger}[\Phi]\,F_{\ell}[\Phi] \}$ corresponding to the Kraus operators $\{F_{\ell}[\Phi]\}$ of each of the photon-added amplifier, attenuator, and phase-conjugation channels are all diagonal in the Fock basis for every $n\geq 0$. 
\label{diag}
\end{proposition}
As mentioned earlier, the above proposition has an information-theoretic consequence with respect to environment assisted communication that we now state as a Theorem. 

\begin{theorem}
For  the photon-added attenuator, amplifier and phase conjugation channels there exists an environment-assisted error correction scheme that restores classical information encoded in the Fock basis for every value $n\geq 0 $ of photon addition. 
\end{theorem}
\proof Consider the following scheme. Let $\{\ket{n_j}\}_{j=0}^{\infty}$ denote the Fock basis with $j$ representing the encoding of a classical message taken from a given message set. Let $\Phi$ denote 
either of the photon-added amplifier, attenuator, or phase-conjugation channels, with Kraus operators $\{ F_{y} \} $. Let $\ket{n_x}\bra{n_x}$ be a particular input state to $\Phi$ encoded with classical message $x$. The output operator conditioned on classical feedback from the environment, say $y$, i.e., conditioned on the action of a particular Kraus operator $F_{y}$, is given by $F_{y} \ket{n_x}\bra{n_x} F_{y}^{\dagger}$. 

We then construct the recovery map (completely positive and trace preserving) as 
\begin{align}
R_{y}(\hr{}) = \sum_{j}  \ket{n_j}\bra{\phi_j} \,\hr{}\,\ket{\phi_j} \bra{n_j} + E_{y} \hr{} E_{y},
\end{align}
where $\ket{\phi_j}$ denotes the normalized state  
\begin{align}\ket{\phi_j} =  (||F_{y}\ket{n_j}||)^{-1}\, F_{y}\ket{n_j}\end{align} for all $||F_{y}\ket{n_j}|| \neq 0$ and $E_{y}$ denotes the projection onto the orthogonal space corresponding to  $ {\rm Span} \,\{ F_{y}\ket{n_j},\, \forall \,j\}$.   Then conditioned on a particular $y$  we have that 
\begin{align}
&R_{y} ( F_{y} \ket{n_x}\bra{n_x} F_{y}^{\dagger} ) \nonumber\\ &=   \sum_{j}  \ket{n_j}  \, \frac{\bra{n_j} F_{y}^{\dagger} F_{y} \ket{n_x}}{(||F_{y}\ket{n_j}||)}
\frac{ \bra{n_x} F_{y}^{\dagger} F_{y}\ket{n_j}}{(||F_{y}\ket{n_j}||)} \, \bra{n_j} \nonumber\\
& ~~~~~~~~+ E_{y}   (F_{y} \ket{n_x}\bra{n_x} F_{y}^{\dagger}) E_{y}.
\end{align}
By Proposition \ref{diag} we have that the above equation reduces to 
\begin{align}
R_{y} ( F_{y} \ket{n_x}\bra{n_x} F_{y}^{\dagger} ) = (\bra{n_x} F_{y}^{\dagger} F_{y}\ket{n_x}) \ket{n_x}\bra{n_x}.
\end{align}
Therefore the action of the corrected channel $\Phi_{\rm corr} = \sum_y R_y \circ F_y$ considering all the Kraus operators (or equivalently labels $y$) on any basis vector $\ket{n_x}$ is then given by 
\begin{align}
\Phi_{\rm corr}(\ket{n_x}\bra{n_x}) &= \sum_{y} R_{y} ( F_{y} \ket{n_x}\bra{n_x} F_{y}^{\dagger} ) \nonumber\\
&=\bra{n_x}  \left [  \sum_{y}  F_{y}^{\dagger} F_{y} \right ] \ket{n_x} \ket{n_x}\bra{n_x} \nonumber \\&=  \ket{n_x}\bra{n_x}.
\end{align}
So  we have that every vector  $\{\ket{n}\}_{0}^{\infty}$ in the Fock basis that is input to $\Phi$ can be  restored by a correction map applied to the output with the help of classical feedback from the environment. In other words, the fact that the positive operators $\{W_{\ell} \}_{0}^{\infty}$ are all diagonal in the Fock basis guarantees that the output states conditioned on a Kraus operator are orthogonal for input Fock states. This implies that these outputs can be reliably distinguished and hence useful for transmission of classical information encoded in them. $\blacksquare$

\subsection{Connection between photon-added channels and noisy Gaussian channels \label{g-ng}}
For this subsection we  introduce some notation. Consider a single-mode Bosonic Gaussian channel represented by the matrix pair $(X,Y)$, $X$ and $Y$ real, and $Y \geq 0$. Its action at the level of the characteristic function  
$\chi(\bs{\xi};\hr{})$ with $\bs{\xi} \in \mathbb{R}^{2}$ is given by~\cite{caruso06,kraus10}
\begin{align}
\chi(\bs{\xi};\hr{}^{\rm out}) = \chi(X\bs{\xi};\hr{}^{\rm in})\,\exp\left[-\frac{1}{2}\bs{\xi}^T\,Y\,\bs{\xi} \right].
\end{align}

Let the noise matrix $Y$ be decomposed as $Y_{\rm QL} + \widetilde{Y}$, where $Y_{\rm QL}$ that stands for quantum-limited is the minimum amount of noise as required by the complete positivity condition for the given $X$, and $\widetilde{Y}$ stands for the additional noise in the channel over and above the amount that makes the channel quantum-limited. For a consistent notation we denote by $\Phi(X,\widetilde{Y};n)$ a $n$ photon added noisy Gaussian channel represented by the matrix pair $(X,Y_{\rm QL} + \widetilde{Y})$, and we suppress the $Y_{\rm QL}$ term and denote only the extra noise $\widetilde{Y}$. We will explicitly consider photon-added noisy Gaussian channels in Section \ref{noisypa}. Therefore, a photon-added quantum-limited channel as considered in earlier sections will be denoted by $\Phi(X,0;n)$ or simply $\Phi(X;n)$.  

We now present a straightforward connection between photon-added Gaussian channels and noisy Gaussian channels. We first begin with the following fact that every noisy Gaussian channel can be obtained by using a thermal state in the Stinespring representation of these noisy channels~\cite{caruso06,kraus10}. In other words we have that any noisy Gaussian channel $(X,Y_{\rm QL} + Y(\bar{n}))$, with additional noise denoted by $Y(\bar{n})$ can be written as 
\begin{align}
\Phi(X,Y(\bar{n});0) =  {\rm Tr}_E\, \left[ U_{AE}\,(\hr{A} \otimes \hr{\rm th}(\bar{n}))\,U_{AE}^{\dagger} \right],
\label{noisy}
\end{align}
where $\hr{\rm th}(\bar{n})$ is the thermal state
\begin{align}
\hr{\rm th}(\bar{n}) =  \sum_{n=0}^{\infty} p_n(\bar{n}) \,|n\rangle \langle n|, \,\,p_n(\bar{n}) =  \frac{\bar{n}^{n}}{(1+\bar{n})^{n+1}}, 
\label{rhoth}
\end{align}
and $\bar{n} = \langle \hat{a}^{\dagger} \hat{a} \rangle_{\hr{\rm th}}$ is the average photon number in the state $\hr{\rm th}(\bar{n})$. By  Eq. \eqref{e8} we have that the amount of additional noise in the Gaussian channel $\Phi(\kappa;0)$ is $Y(\bar{n}) = 2\bar{n} \,Y_{\rm QL}$ (with the quantum-limited channel corresponding to $\bar{n}=0$, or vacuum state in the environment).
  
Substituting the Fock basis representation of  Eq. \eqref{rhoth} in  Eq. \eqref{noisy}  we have that 
\begin{align}
&\Phi(X,Y(\bar{n});0) \nonumber\\
&=  {\rm Tr}_E\, \left[ U_{AE}\,(\hr{A} \otimes  (1-a) a^n \ket{n}\bra{n})\,U_{AE}^{\dagger} \right] \nonumber\\
&=  (1-a) \sum_{n=0}^{\infty} a^n \, {\rm Tr}_E\, \left[ U_{AE}\,(\hr{A} \otimes \ket{n}\bra{n} )\,U_{AE}^{\dagger} \right] \nonumber\\
&=  \sum_{n=0}^{\infty} p_n(\bar{n}) \Phi(X;n), ~ a =  \frac{\bar{n}}{1+\bar{n}}, 
\end{align}
where $\Phi(X;n)$ are the photon-added Gaussian (quantum-limited) channels. To summarize we have 
\begin{align}
\Phi(X,Y(\bar{n});0) = \sum_{n=0}^{\infty} p_n(\bar{n}) ~\Phi(X;n) 
\label{thermal}
\end{align}
which is expressed as the following proposition connecting noisy Gaussian channels and the non-Gaussian photon-added channels. 

\begin{proposition}
The noisy  attenuator, amplifier, and phase conjugation channels can be decomposed into a convex mixture of their corresponding photon-added channels as given in  Eq. \eqref{thermal}. 
\label{noisydeco}
\end{proposition}
Thermal states can be thought of as `noisy' Gaussian states and these can be decomposed in the Fock basis with a suitable geometric distribution as given in  Eq. \eqref{rhoth}. Likewise Proposition
 \ref{noisydeco} provides a representation for the channel analogue wherein a noisy Gaussian channel can be decomposed into a convex mixture of photon-added Gaussian channels also with a geometric distribution, where photon-added Gaussian channels take on the role of Fock states. So we see how these photon-added channels are fundamentally related to the ubiquitous noisy Gaussian channels.

\subsection{Photon addition to noisy channels: \protect\\ a complete framework \label{noisypa}}
\begin{figure}
\centering
\includegraphics[width=\columnwidth]{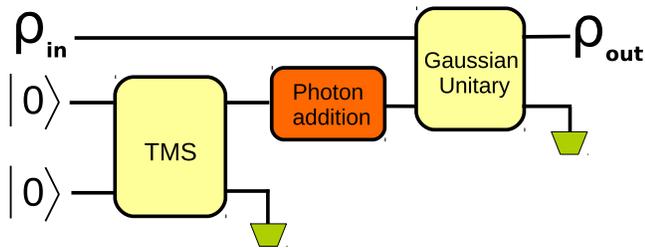}
\caption{Schematic for photon-added noisy Gaussian channels with a vacuum environment state. In the absence of photon addition we obtain the underlying noisy Gaussian channel with noise parameter tuned by the squeeze parameter of the two-mode squeeze (TMS) operator.  \label{fig10}}
\end{figure}

Up to now we have considered photon-added quantum-limited Gaussian channels. In this subsection we generalize the scheme to incorporate photon addition to noisy Gaussian channels in a systematic way. To achieve this, we start from the unitary dilation of noisy Gaussian channels. There are two possible implementations for photon addition that one can take depending on the choice of the environment state being pure or mixed. For the first case of pure vacuum state we have the following proposition \cite{kraus10}.
\begin{proposition}
Noisy attenuator, amplifier, and phase-conjugation channels of a single-mode can be realized with a two-mode vacuum environment state followed by a suitable two-mode squeeze transformation before the application of the channel unitary. 
\end{proposition}
With this realization of noisy Gaussian channels, one can then apply a photon addition to obtain the photon-added noisy Gaussian channels as schematically depicted in Fig. \ref{fig10}.

\begin{figure}
\centering
\includegraphics[width=\columnwidth]{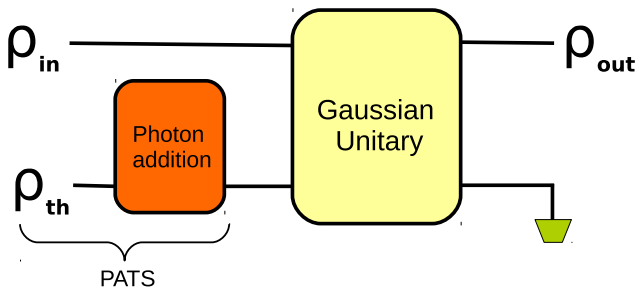}
\caption{Schematic for photon-added noisy Gaussian channels with a mixed environment state. Here PATS stands for photon-added thermal state.  \label{fig11}}
\end{figure}

Equivalently, we can choose to trace out the second environment mode resulting in a mixed thermal state for the first environment mode to which one applies photon addition. We recall that this scheme was mentioned in Sec. \ref{g-ng} to connect noisy Gaussian channels and photon-added Gaussian channels. Then by photon addition we get a photon-added thermal state (PATS) \cite{pats1,pats2} for the environment as depicted in Fig. \ref{fig11}. With respect to the resulting photon-added channel, the two schemes are identical. 

To summarize we provided a uniform framework that includes four special cases. We focused on the first scheme with a pure environment state, with equivalent statements that can be easily inferred for the case of a mixed environment state. First, in the absence of both the photon addition and the two-mode squeezing (TMS) we have the representation for a quantum-limited Gaussian channel. Second, to obtain the noisy version of the Gaussian channel one can tune the squeezing parameter of the TMS and with no photon addition. Third, to obtain photon addition to the quantum-limited Gaussian channels we switch off the TMS. Finally, with all elements in place we have photon addition to noisy Gaussian channels. 
We thus have a consistent description of photon addition as in the earlier cases of photon addition to quantum-limited Gaussian channels. Also in this way the role of each component is made transparent. We will consider the properties and applications of photon-added noisy Gaussian channels elsewhere.  

\section{Conclusions}
We presented a systematic framework to study linear Bosonic non-Gaussian channels of continuous-variable systems. We focused on a special class of channels we call photon-added Gaussian channels. Photon-added Gaussian channels are realized by extending quantum-limited Gaussian channels with photon addition applied to the environment state in the Stinespring dilation of these channels. The resulting channels are linear and non-Gaussian in nature. We considered the case of photon-added attenuator, amplifier, and phase conjugator as our main examples as these channels can be realized using either a beamsplitter or a two-mode squeeze operator, both of which are very important and practically realized with current quantum-optical technology. We first obtained the operator-sum representation of the photon-added channels and studied various implications. 

Our most interesting application is to the case of activation of quantum-optical nonclassicality at outputs of channels that would otherwise output only classical states. This was achieved through the use of photon addition in the ancilla modes. Further, we also report activation of the quantum and private capacities of these channels. This has potentially far-reaching applications in quantum-optical communications. We emphasize that no operation on the input state could have achieved this phenomena of activation of the said quantities, giving a clear distinction and use of our scheme of photon addition to the ancilla states.

The action of the channels on Fock states revealed the ranges of the Fock support for the output states corresponding to any input states. We also see that the photon-added  phase-conjugation channel is complementary to the photon-added amplifier channel, and the photon-added attenuator channel is complementary to another photon-added attenuator channel with different attenuation parameter, leading to an immediate implication on the minimum output entropy of these channels. From the structure of the Kraus operators  we also showed that all photon-added Gaussian channels that we considered allow for environment-assisted classical information transmission through them.

The photon-added amplifier, attenuator, and phase-conjugation channels are also coherence non-generating in the Fock basis when acting on incoherent states. In respect to the resource theory of non-Gaussianity, it was also noted that input Gaussian states are mapped to non-Gaussian states at the output, therefore resource generating, in contrast to their behaviour with respect to Fock-basis coherence. 

We then provided a decomposition of noisy Gaussian channels in terms of their respective photon-added quantum-limited channels analogous to the Fock basis representation of a thermal state, and thereby providing further motivation for studying these non-Gaussian channels as they are fundamentally related to noisy Gaussian channels.

Furthermore, our method allows for tuning between the Gaussian and non-Gaussian regime in the space of channels  through photon addition where $n$ plays the role of the tuning parameter, with $n=0$ corresponding to a Gaussian channel and $n>0$ corresponding to a non-Gaussian channel.

The present study is one approach contributing to the systematic study of non-Gaussian operations that have not only proved advantageous for many quantum information protocols but are also necessary due to many Gaussian no-go theorems as mentioned in the Introduction. In light of recent interest with quantum resource theories, it would be useful to quantify the resource-theoretic properties of the photon-added channels for resources like quantum-optical nonclassicality and non-Gaussianity \cite{paris1,paris2,simon1,cerfng}, whose resource theories are still important open problems in continuous-variable quantum information theory.

The photon-added channels considered in this article are experimentally realizable as gadgets with potential use in quantum protocols since they only use simple Gaussian unitary elements along with photon addition. We believe that there are many applications of the present work in light of the increasing and indispensable use of non-Gaussian resources in continuous-variable quantum computing, cryptography, and communications tasks.

\subsection*{Acknowledgements}
We acknowledge useful discussions with Nicolas Cerf, Raul Garc\'{i}a-Patr\'{o}n
and Dong Yang. The authors are supported by the ERC, Advanced Grant  ``IRQUAT'', 
contract no.~ERC-2010-AdG-267386, the Spanish MINECO, projects no.~FIS2013-40627-P
and FIS2016-86681-P, with the support of FEDER funds, 
and the Generalitat de Catalunya CIRIT, project no.~2014-SGR-966.

\section{Appendix}
\subsection{Structure of quadratic operators $\{F^{\dagger}_iF_j \}$ \label{quad} and extremality}
For a given channel $\Phi$ there are many useful properties that can be inferred from the structure of the corresponding quadratic operators $\{F^{\dagger}_i[\Phi] F_j[\Phi]\}_{ij}$, where $\{F_j[\Phi] \}_j $ are the Kraus operators of $\Phi$. Extremality is one such example and another application to environment-assisted communication was presented earlier in Sec. \ref{envasec}.

The set of all channels on a single-mode system form a convex set. Extremal channels are those that lie on the boundary of this convex set and cannot be expressed as a convex combination of other channels. Starting from the structure of the Kraus operators, we state the following conjecture in respect of the photon-added channels and extremality. 
\begin{conjecture}
The photon-added amplifier, attenuator, and, phase conjugation channels are extremal  in the set of all completely positive trace-preserving maps for every $n>0$.   
\label{extremus}
\end{conjecture}
The conjecture was shown to hold true for the case $n=0$, i.e., the Gaussian channel case, in Ref. \cite{kraus10} using a theorem by Choi~\cite{choi,tsui} that tells us that a channel $\Phi$ with Kraus operators $\{F_{\ell }[\Phi] \}_{\ell}$    is extremal if and only if the set of associated matrices $\{ W[\Phi]_{\ell \ell^{\prime}} \}_{\ell,\ell^{\prime}}$ defined as
\begin{align}
W[\Phi]_{\ell \ell^{\prime}} := F_{\ell}^{\dagger}[\Phi]\, F_{\ell^{\prime}}[\Phi], ~~ \ell,\ell^{\prime}=0,1,\cdots,\infty 
\end{align}
are linearly independent.

We now compute the quadratic operators for each of the channels beginning with the photon-added phase conjugation channel. 
\begin{align}
&W[{\cal C}(\kappa;n)]_{\ell \ell^{\prime}} \nonumber\\
&= d^{2}_{\kappa}  \sum_{n_1 =  0 }^{n+\ell}\sum_{m_1 =  0 }^{n+\ell^{\prime}} \sqrt{\frac{n!\ell!}{(\ell + n-n_1)!n_1!}}  \sqrt{\frac{n!\ell^{\prime}!}{(\ell^{\prime} + n-m_1)!m_1!}}~\nonumber\\ & \times g_3(n,n_1,\ell+n-n_1;\kappa) ~ g_3(n,m_1,\ell^{\prime}+n-m_1;\kappa) \nonumber \\
&~\times \braket{\ell + n - n_1}{\ell^{\prime}+n-m_1} \ket{n_1} \bra{m_1} \nonumber\\
&=d^{2}_{\kappa}  \sum_{n_1 =  \max \{0,\,\ell - \ell^{\prime}\} }^{n+\ell} \sqrt{\frac{n!^2\ell!\ell^{\prime}! (\ell^{\prime}-\ell+n_1)!^{-1}}{n_1!(\ell-n+n_1)!^2}} \nonumber\\ &~~\times 
 \,g_3(n,\ell^{\prime}-\ell+n_1,\ell+n-n_1;\kappa) \nonumber\\ & ~~~ \times  \,g_3(n,n_1,\ell+n-n_1;\kappa)\,\ket{n_1} \bra{\ell^{\prime}-\ell+n_1}.
\label{wc}
\end{align}

We next consider the quadratic operators $\{W[{\cal B}(\kappa;n)]_{\ell \ell^{\prime}}\}$ for the attenuator channel. We have by  Eq. \eqref{bosr} that 
\begin{align}
&W[{\cal B}(\kappa;n)]_{\ell \ell^{\prime}} 
= B_{\ell}^{\dagger}(\kappa;n) B_{\ell^{\prime}}(\kappa;n) \nonumber\\ 
&= \sum_{n_1 = \max \{0, \ell -n\}}^{\infty} \sum_{m_1 = \max \{0, \ell -n\}}^{\infty}   \sqrt{\frac{(n_1+n-\ell)!\ell!}{n_1!n!}}\nonumber\\ & \times  \sqrt{\frac{(m_1+n-\ell^{\prime})!\ell^{\prime}!}{m_1!n!}}   
 g_2(n,\ell,n_1;\kappa) g_2(n,\ell^{\prime},m_1;\kappa)  \nonumber\\ & ~~~~~\times \braket{n_1 + n-\ell}{m_1 + n-\ell^{\prime}} \ket{n_1} \bra{m_1}\nonumber\\
&= \sum_{n_1 = \max\{0,\ell-n,\ell - \ell^{\prime}\}}^{\infty}
\sqrt{\frac{(n_1+n-\ell)!^2\ell!\ell^{\prime}!}{n_1!n!^2(n_1+\ell^{\prime} - \ell)!}} \, g_2(n,\ell,n_1;\kappa)
 \nonumber\\
& ~~~~~~\times  \,g_2(n,\ell^{\prime},n_1+\ell^{\prime} - \ell;\kappa) \ket{n_1} \bra{n_1+\ell^{\prime} - \ell}.
\label{wb}
\end{align}

As the final case the operators quadratic in the Kraus operators for the photon-added amplifier channel are given by Eq. \eqref{aosr} to be 
\begin{align}
&W[{\cal A}(\kappa;n)]_{\ell \ell^{\prime}} = A_{\ell}^{\dagger}(\kappa;n) A_{\ell^{\prime}}(\kappa;n) \nonumber\\
&=
\kappa^{-2}\sum_{n_1 =\max \{ 0,n-\ell\} }^{\infty} \sum_{m_1 = \max \{ 0,n-\ell^{\prime}\} }^{\infty}  \sqrt{\frac{n_1!\ell!}{(\ell -n + n_1)!n!}}  \nonumber\\  & \times \,g_1(n,n_1,\ell-n +n_1;\kappa)\, g_1(n,m_1,\ell^{\prime}-n+m_1;\kappa)  \nonumber\\
&\times  \sqrt{\frac{m_1!\ell^{\prime}!}{(\ell^{\prime} -n + m_1)!n!}}\,\braket{n_1 + \ell - n}{m_1 +\ell^{\prime}-n} \ket{n_1}  \bra{m_1}. 
\end{align}
Upon simplifying the above equation we evaluate $W[{\cal A}(\kappa;n)]_{\ell \ell^{\prime}}$ to be 
\begin{align}
&\kappa^{-2} \sum_{n_1 =\max \{ 0,n-\ell,\ell^{\prime}-\ell\}}^{\infty} 
\sqrt{\frac{n_1!\ell!\ell^{\prime}!(n_1+\ell -\ell^{\prime})}{(\ell -n + n_1)!^2n!^2}} \,\nonumber\\
&\times \, g_1(n,n_1 + \ell - \ell^{\prime},\ell -n +n_1;\kappa)\,\nonumber\\
&\times \,g_1(n,n_1,\ell-n+n_1;\kappa)~\ket{n_1}\bra{n_1 + \ell - \ell^{\prime}}.
\label{wa}
\end{align}

We see that there are three natural branches corresponding to $\ell < \ell^{\prime}$, $\ell > \ell^{\prime}$ and $\ell = \ell^{\prime}$ in which the corresponding quadratic operators are mutually linearly independent. Proving that the quadratic operators within each branch are also linearly independent would settle Conjecture \ref{extremus} by the above mentioned Theorem of Choi. 

\begin{remark}
We  reiterate that demonstrating  a channel to be extremal implies that the channel is fundamentally one that cannot be decomposed into a convex combinations of other constituent channels. 
\end{remark}

\subsection{Proof of Proposition \ref{compcha} \label{compli}}

We begin at the level of the Fock basis matrix elements of the associated unitary operators in their respective channel dilations. We first prove the following relation that 
\begin{align}
U[{\cal C}(\sqrt{\kappa^2-1)}] = U[{\cal A}(\kappa)] \circ U[{\cal B}(0)] \circ U_{lo}, 
\end{align} 
where $U_{lo}$ is a product of local unitary operators and is given by 
\begin{align}
\bra{m_1,m_2} U_{lo} \ket{n_1,n_2} = (-1)^{n_2} \delta_{m_1,n_1} \delta_{m_2,n_2}.
\end{align}
Equivalently, we have that 
\begin{align}
U_{lo} = 1\!\!1 \otimes \sum_{n_2} (-1)^{n_2} \proj{n_2}. 
\end{align}
Its role is to compensate for the phases.  Furthermore, we have by Eq. \eqref{buni} that 
\begin{align}
\bra{m_1,m_2} U[{\cal B}(0)]\ket{n_1,n_2}   =  (-1)^{n_2} \delta_{m_1,n_2} \delta_{m_2,n_1}.
\end{align}
We have the product of $U[{\cal B}(0)] \circ U_{lo}$  to be the flip operator $U_{\rm flip}$ which is given by
\begin{align}
\bra{m_1,m_2} U_{\rm flip}\ket{n_1,n_2} = \delta_{m_1,n_2} \delta_{m_2,n_1}.
\end{align}
By Eq. \eqref{auni} we see that $U[{\cal A}(\kappa)] \circ U_{\rm flip}$ is given by 
\begin{align*}
&\bra{m_1,m_2} U[{\cal A}(\kappa)] \circ U_{\rm flip} \ket{n_1,n_2} 
= \bra{m_1,m_2} U[{\cal A}(\kappa)] \ket{n_2,n_1} \nonumber\\ 
&=  {\kappa}^{-1} \sqrt{\frac{n_2!m_2!}{m_1!n_1!}} \sum_{j=0}^{n_1} \sum_{r=0}^{m_1}   \binom{n_1}{r} \binom{m_1}{r} \, (-1)^{j} \,(\kappa^{-1})^{n_1+r-j} \nonumber\\ & \times \,(\sqrt{1-\kappa^{-2}})^{m_1+j-r}  \delta_{n_2,r+j} \delta_{m_2,n_1+m_1-r-j}.
\end{align*} 
We substitute $\kappa = \sqrt{1 + t^{2}}$ which can always be done since $\kappa \geq 1$. We have that $\bra{m_1,m_2} U[{\cal A}(t)] \ket{n_2,n_1} $ is given by 
\begin{align}
&({\sqrt{1 + t^{2}}})^{-1}  \sqrt{\frac{n_2!m_2!}{m_1!n_1!}}  \sum_{j=0}^{n_1} \sum_{r=0}^{m_1}  \binom{m_1}{r} (\sqrt{1 + t^{2}})^{-(n_1+r-j)}   \nonumber\\
& \times  \binom{n_1}{r} (-1)^{j} (\sqrt{1+t^{-2}})^{-(m_1+j-r)} \delta_{n_2,r+j} \delta_{m_2,n_1+m_1-r-j}\nonumber\\
&= T[{\cal C}(t)]^{m_1,m_2}_{n_1,n_2}, 
\end{align}
where the last line followed from Eq. \eqref{cuni}. Hence, we recover the matrix elements of the phase conjugation unitary. It is then a simple consequence that the $n$ photon-added phase conjugation ${\cal C}(\sqrt{\kappa^2 -1};n)$ is complementary to the $n$ photon-added amplifier ${\cal A}(\kappa;n)$ for each $n \geq 0$. 

The proof of the second part of the theorem regarding attenuators also follows in a similar way to the above proof since 
\begin{align}
U[{\cal B}(\sqrt{1-\kappa^2})] = U_{\rm flip} \circ U[{\cal B}(\kappa)], 
\end{align}
as can be seen by Eq. \eqref{buni}. Hence the photon-added attenuator ${\cal B}(\kappa;n)$ is complementary to another photon-added attenuator ${\cal B}(\sqrt{1-\kappa^2};n)$ for all $n \geq 0$. The relation between the unitary operators of the photon-added amplifier and the photon-added phase conjugation, and the relation between the two beamsplitters is represented in Fig. \ref{fig9}. \hfill $\blacksquare$


\begin{thebibliography}{99}
\bibitem{fock1}  M. Hofheinz, E. M. Weig, M. Ansmann, R. C. Bialczak, E. Lucero, M. Neeley, A. D. O'Connell, H. Wang, J. M. Martinis, and A. N. Cleland, Generation of Fock states in a superconducting quantum circuit, Nature {\bf 454}, 310 (2008).  
\bibitem{fock2}  C. Sayrin et. al., Real-time quantum feedback prepares and stabilizes photon number states, Nature {\bf 477}, 73 (2008). 
\bibitem{motes}  K. R. Motes,  R. L. Mann, J. P. Olson, N. M. Studer, E. A. Bergeron, A. Gilchrist, J. P. Dowling, D. W. Berry, and P. P. Rohde, Efficient recycling strategies for preparing large Fock states from single-photon sources:  Applications to quantum metrology, \pra {\bf 94}, 012344 (2016).
\bibitem{noon}  I. Afek,  O. Ambar, and Y. Silberberg, High-NOON states by mixing quantum and classical light, Science {\bf 328}, 879 (2010). 
\bibitem{suzuki} M. Sasaki and M. Suzuki, Multimode theory of measurement-induced non-Gaussian operation on wideband squeezed light: Analytical formula, Phys. Rev. A {\bf 73}, 043807 (2006). 
\bibitem{cat} A. Ourjoumtsev, H. Jeong, R. Tualle-Brouri, and P. Grangier,   Generation of optical `Schr\"{o}dinger cats' from photon number states, Nature {\bf 448}, 784 (2007). 
\bibitem{agarwal}  G. S. Agarwal and K. Tara, Nonclassical properties of states generated by the excitations on a coherent state, Phys. Rev. A {\bf 43}, 492 (1991).
\bibitem{bellini04} A. Zavatta, S. Viciani, and M. Bellini, Quantum-to-Classical Transition with Single-Photon-Added Coherent States of Light, Science {\bf 306}, 660 (2004).
\bibitem{pats1}  G. S. Agarwal and K. Tara,  Nonclassical character of states exhibiting no squeezing or sub-Poissonian statistics, Phys. Rev. A {\bf 46}, 485 (1992). 
\bibitem{pats2}  T. Keisel,  W. Vogel, V. Parigi,  A. Zavatta, and M. Bellini, Experimental determination of a nonclassical Glauber-Sudarshan P function, Phys. Rev. A {\bf 78}, 021804(R) (2008). 
\bibitem{grangier04}  J. Wenger, R. Tualle-Brouri, and P. Grangier,  Non-Gaussian Statistics from Individual Pulses of Squeezed Light, Phys. Rev. Lett. {\bf 92}, 153601 (2004).
\bibitem{polzik06} J. S. Neergaard-Nielsen, B. M. Nielsen, C. Hettich,  K. M\o lmer, and E. S. Polzik, Generation of a Superposition of Odd Photon Number States for Quantum Information Networks, Phys. Rev. Lett. {\bf 97}, 083604 (2006).  
\bibitem{grangier12} J. Stanojevic, V. Parigi, E. Bimbard, A. Ourjoumtsev,  P. Pillet, and P. Grangier, Generating non-Gaussian states using collisions between Rydberg polaritons, Phys. Rev. A  {\bf 86}, 021403(R)   (2012).
\bibitem{bellini07}  V. Parigi, A. Zavatta, M. S. Kim, and M. Bellini, Probing quantum commutation rules by addition and subtraction of single photons to/from a light field,  Science {\bf 317}, 1890 (2007).
\bibitem{fiurasek09} J. Fiur\'{a}\v{s}ek, Engineering quantum operations on traveling light beams by multiple photon addition and subtraction, Phys. Rev. A {\bf 80}, 052822 (2009).
\bibitem{chefles}  A. Kitagawa, M. Takeoka, M. Sasaki, and A. Chefles, Entanglement evaluation of non-Gaussian states generated by photon subtraction from squeezed states, Phys. Rev. A {\bf 73}, 042310 (2006).
\bibitem{psub}  N. Namekata, Y. Takahashi, G. Fujii, D. Fukuda, S. Kurimura, and S. Inoue, Non-Gaussian operation based on photon subtraction using a photon-number-resolving detector at a telecommunications wavelength, Nat. Photonics {\bf 4}, 655 (2010).  
\bibitem{sasaki07} K. Wakui, H. Takahashi, A. Furusawa, and M. Sasaki, Photon subtracted squeezed states generated with periodically poled $KTiOPO_4$, Opt. Exp. {\bf 15}, 3568 (2007).
\bibitem{pcoun} C. Guerlin, J. Bernu, S. Del\'{e}glise, C. Sayrin, S. Gleyzes, S. Kuhr, M. Brune, J. Raimond, and S. Haroche, Progressive field-state collapse and quantum non-demolition photon counting, Nature {\bf 448}, 889 (2007).  
\bibitem{miyata}  K. Miyata, H. Ogawa, P. Marek, R. Filip, H. Yonezawa, J. Yoshikawa, and A. Furusawa, Implementation of a quantum cubic gate by an adaptive non-Gaussian measurement, Phys. Rev. A {\bf 93}, 022301 (2016).
\bibitem{kerr}  T. Tyc and N. Korolkova, Highly non-Gaussian states created via cross-Kerr nonlinearity,  New J. Phys. {\bf 10}, 023041  (2008).
\bibitem{nogo1} J. Eisert, S. Scheel, and M. B. Plenio, Distilling Gaussian states with Gaussian operations is impossible, Phys. Rev. Lett. {\bf 89}, 137903 (2002).
\bibitem{nogo2} J. Fiur\'{a}\v{s}ek,  Gaussian transformations and distillation of entangled Gaussian states, \prl {\bf 89}, 137904 (2002).
\bibitem{nogo3}  G. Giedke and J. I. Cirac,  Characterization of Gaussian operations and distillation of Gaussian states,  \pra {\bf 66}, 032316  (2002).
\bibitem{guha14} R. Namiki, O. Gittsovich,  S. Guha, and N. L\"{u}tkenhaus, Gaussian-only regenerative stations cannot act as quantum repeaters, \pra  {\bf 90}, 062316 (2014).
\bibitem{wolf}  N. J. Cerf, O. Kr\"{u}ger, P. Navez, R. F. Werner, and M. M. Wolf, Non-Gaussian Cloning of Quantum Coherent States is Optimal, \prl {\bf 95}, 070501 (2005). 
\bibitem{cerf09}  J. Niset, J.  Fiur\'{a}\v{s}ek,  and N. J. Cerf, No-Go Theorem for Gaussian Quantum Error Correction, \prl {\bf 102}, 120501 (2009). 
\bibitem{cerf10}  L. Magnin, F. Magniez, A. Leverrier, and N. J. Cerf, Strong no-go theorem for Gaussian quantum bit commitment, \pra  {\bf 81}, 010302(R) (2010). 
\bibitem{eisert10}  M. Ohliger, K. Kieling, and J. Eisert, Limitations of quantum computing with Gaussian cluster states, \pra {\bf 82}, 042336 (2010). 
\bibitem{lloyd}  S. Lloyd and S. L. Braunstein, Quantum Computation over Continuous Variables, \prl  {\bf 82},  1784 (1999).
\bibitem{sanders}  S. D. Bartlett and B. C. Sanders, Universal continuous-variable quantum computation: Requirement of optical nonlinearity for photon counting, \pra {\bf 65}, 042304 (2002). 
\bibitem{nielsen}  N. C. Menicucci, P. van Loock, M. Gu, T. C. Ralph, and M. A. Nielsen, Universal Quantum Computation with Continuous-Variable Cluster States, \prl {\bf 97}, 110501 (2006). 
\bibitem{adesso} G. Adesso,  F. Dell'Anno,  S. De Siena, F. Illuminati, and  L. A. M. Souza, Optimal estimation of losses at the ultimate quantum limit with non-Gaussian states, \pra {\bf 79}, 040305(R) (2009).
\bibitem{grangier07}  A. Ourjoumtsev, A. Dantan, R. Tualle-Brouri, and  P. Grangier, Increasing Entanglement between Gaussian States by Coherent Photon Subtraction, \prl {\bf 98}, 030502 (2007).
\bibitem{sasaki}  H. Yakahashi, J. S. Neergaard-Nielsen, M. Takeuchi, M. Takeoka, K. Hayasaka,  A. Furusawa, and M. Sasaki, Entanglement distillation from Gaussian input states, Nat. Photonics  {\bf 4}, 178 (2010).
\bibitem{robust} K. K. Sabapathy, J. S. Ivan, and R. Simon, Robustness of Non-Gaussian Entanglement against Noisy Amplifier and Attenuator Environments, \prl {\bf 107}, 130501  (2011).
\bibitem{raul12}  C. Navarrete-Benlloch,  R. Garc\'{i}a-Patr\'{o}n, J. H. Shapiro, and N. J. Cerf, Enhancing quantum entanglement by photon addition and subtraction, \pra {\bf 86}, 012328, (2012). 
\bibitem{opat}  T. Opatrn\'{y}, G. Kurizki, and  D.-G. Welsch, Improvement on teleportation of continuous variables by photon subtraction via conditional measurement, \pra {\bf 61}, 032302 (2000).
\bibitem{bonifacio} S. Olivares, M. G. A. Paris, and R. Bonifacio, Teleportation improvement by inconclusive photon subtraction, \pra {\bf 67}, 032314 (2003).  
\bibitem{mista06} L. Mi\^{s}ta, Jr., Minimal disturbance measurement for coherent states is non-Gaussian, Phys. Rev. A 73, 032335 (2006). 
\bibitem{anno07} F. Dell'Anno, S. De Siena, L. Albano, and F. Illuminati, Continuous-variable quantum teleportation with non-Gaussian resources, \pra {\bf 76}, 022301 (2007). 
\bibitem{barry} S. Ghose and B. C. Sanders, Non-Gaussian ancilla states for continuous variable quantum computation via Gaussian maps, J. Mod. Opt. {\bf 54}, 855 (2005).
\bibitem{smolin} B. M. Terhal, I. L. Chuang, D. P. DiVincenzo,  M. Grassl,  and J. A. Smolin, Simulating quantum operations with mixed environments, \pra {\bf 60}, 881 (1999).
\bibitem{sid1}  S. Karumanchi, S. Mancini, A. Winter, and D. Yang, Quantum Channel Capacities with Passive Environment Assistance, IEEE Trans. Inf. Theory {\bf 62}, 1733 (2016).
\bibitem{sid2}  S. Karumanchi, S. Mancini, A. Winter, and D. Yang, Classical capacities of quantum channels with environment assistance, arXiv:1602.02036 [quant-ph].  
\bibitem{andi.env} J. A. Smolin, F. Verstraete, and A. Winter, Entanglement of assistance and multipartite state distillation, \pra {\bf 72}, 052317 (2005).
\bibitem{grego1}  M. Gregoratti and R. F. Werner,  Quantum lost and found, J. Mod. Opt. {\bf 50} 915 (2003).
\bibitem{grego2}  M. Gregoratti and R. F. Werner, On quantum error-correction by classical feedback in discrete time, J. Math. Phys. {\bf 45}, 2600 (2004).
\bibitem{king} P. Hayden and C. King, Correcting quantum channels by measuring the environment, Quantum Inf. Comput. {\bf 5}, 156 (2005).
\bibitem{kraus} K. Kraus,  States, Effects, and Operations: Fundamental Notions in Quantum Theory, Lecture Notes in Physics Vol. 190 (Springer-Verlag, Berlin) (1983).
\bibitem{nb} J. S. Ivan, K. K. Sabapathy, and R. Simon, Nonclassicality breaking is the same as entanglement breaking for Bosonic Gaussian channels, \pra {\bf 88}, 032302 (2013). 
\bibitem{nbm} K. K. Sabapathy, Quantum-optical channels that output only classical states, \pra {\bf 92}, 052301 (2015).
\bibitem{hw01}  A. S. Holevo and R. F. Werner, Evaluating capacities of Bosonic Gaussian channels, \pra {\bf 63}, 032312 (2001). 
\bibitem{simon00} R. Simon, Peres-Horodecki Separability Criterion for Continuous Variable Systems, \prl {\bf 84}, 2726 (2000). 
\bibitem{kraus10} J. S. Ivan, K. K. Sabapathy, and R. Simon, Operator-sum representation for Bosonic Gaussian channels, \pra  {\bf 84}, 042311 (2011). 
\bibitem{gio17} G. De Palma, D. Trevisan, and V. Giovannetti, Gaussian States Minimize the Output Entropy of the One-Mode Quantum Attenuator, IEEE Trans. Inf. Theory {\bf 63}, 728 (2017).
\bibitem{ntoE1} M. S. Kim, W. Son, V. Bu\^{z}ek, and  P. L. Knight,
Entanglement by a beam splitter: Nonclassicality as a prerequisite for entanglement, \pra {\bf 65}, 032323 (2002).
\bibitem{ntoE2}  J. S. Ivan, S. Chaturvedi, E. Ercolessi, G. Marmo, G. Morandi,
N. Mukunda, and R. Simon, Entanglement and nonclassicality
for multimode radiation-field states, \pra {\bf 83}, 032118 (2011).
\bibitem{ntoE3} Z. Jiang, M. D. Lang, and C. M. Caves, Mixing nonclassical
pure states in a linear-optical network almost always generates modal entanglement, \pra {\bf 88}, 044301 (2013).
\bibitem{roman} R. Schnabel, Squeezed states of light and their applications in laser interferometers, arXiv:1611.03986 [quant-ph].  
\bibitem{ecg} E. C. G. Sudarshan, Equivalence of semiclassical and quantum mechanical descriptions of statistical light beams, \prl {\bf 10}, 277 (1963).
\bibitem{pon} K. K. Sabapathy, Process output nonclassicality and nonclassicality depth of quantum-optical channels, \pra {\bf 93}, 042103 (2016).
\bibitem{lut}  N. L\"{u}tkenhaus and S. M. Barnett, Nonclassical effects in phase space, \pra {\bf 51}, 3340 (1995).
\bibitem{wildebook} M. M. Wilde, Quantum Information theory {\em Cambridge University Press} (2013).
\bibitem{shor04} P. W. Shor, Additivity of the classical capacity of entanglement-breaking quantum channels,  J. Math. Phys. {\bf 43}, 4334  (2002).
\bibitem{holevo08} A. S. Holevo, Entanglement-breaking channels in infinite
dimensions, Probl. Inf. Transm. {\bf 44}, 171 (2008).  
\bibitem{schumi96} B. Schumacher and M. A. Nielsen,  Quantum data processing and error correction, \pra {\bf 54}, 2629 (1996). 
\bibitem{dong} A. Winter and D. Yang, Operational Resource Theory of Coherence, \prl {\bf 116}, 120404 (2016).
\bibitem{plenio}  T. Baumgratz, M. Cramer, and M. B. Plenio, Quantifying Coherence, \prl {\bf 113}, 140401 (2014).
\bibitem{eric} E. Chitambar and G. Gour, Are Incoherent Operations Physically Consistent? A Critical Examination of Incoherent Operations, arXiv:1602.06969 [quant-ph].
\bibitem{raul16} M. G. Jabbour, R. Garc\'{i}a-Patr\'{o}n, and N. J. Cerf, Majorization preservation of Gaussian Bosonic channels, New J. Phys. {\bf 18}, 073047 (2016).
\bibitem{caruso06} F. Caruso,  V. Giovannetti, and A. S. Holevo, One-mode Bosonic Gaussian channels: a full weak-degradability classification, New J. Phys. {\bf 8}, 310 (2006).
\bibitem{choi} M. D. Choi, Completely positive linear maps on complex matrices, Lin. Alg. Appl. {\bf 10}, 285 (1975).
\bibitem{tsui}  S.-K. Tsui, Completely positive module maps and completely positive extreme maps, Proc. Am. Math. Soc. {\bf 124}, 437 (1996).
\bibitem{paris1} M. G. Genoni, M. G. A. Paris, and K. Banaszek, Quantifying the non-Gaussian character of a quantum state by quantum relative entropy, \pra {\bf 78}, 060303(R) (2008).
\bibitem{paris2} M. G. Genoni and M. G. A. Paris, Quantifying non-Gaussianity for quantum information, \pra {\bf 82}, 052341 (2010).  
\bibitem{simon1} J. S. Ivan, M. S. Kumar, and R. Simon, Quantum Inf. Process {\bf 11}, 853 (2012). 
\bibitem{cerfng} A. Mandilara and N. J. Cerf, Quantum uncertainty relation saturated by the eigenstates of the harmonic oscillator, Phy. Rev. A {\bf 86}, 030102(R) (2012).
\end{thebibliography}
\end{document}